\documentclass[manuscript]{aastex}
\usepackage{amsmath}
\usepackage[colorlinks=true,citecolor=blue]{hyperref}
\usepackage{graphicx,color}
\usepackage{times}
\usepackage{amssymb}

\newcommand{\lum}{erg\,s$^{-1}$}
\newcommand{\fermi}{{\it Fermi}}
\newcommand{\nustar}{{\it NuSTAR}}
\newcommand{\xmm}{{\it XMM-Newton}}
\newcommand{\swift}{{\it Swift}}
\newcommand{\phflux}{\mbox{${\rm \, ph \,\, cm^{-2} \, s^{-1}}$}}

\newcommand{\gm}{$\gamma$}

\slugcomment{ApJ accepted}

\shorttitle{High Redshift \nustar~Blazars}
\shortauthors{Paliya et al.}

\begin{document}

\title{Broadband Observations of High Redshift Blazars}

\author{Vaidehi S. Paliya$^{1}$, M. L. Parker$^{2}$, A. C. Fabian$^{2}$, and C. S. Stalin$^{3}$} 
\affil{$^1$Department of Physics and Astronomy, Clemson University, Kinard Lab of Physics, Clemson, SC 29634-0978, USA}
\affil{$^2$Institute of Astronomy, Madingley Road, Cambridge CB3 0HA, UK}
\affil{$^3$Indian Institute of Astrophysics, Block II, Koramangala, Bangalore-560034, India}
\email{vpaliya@g.clemson.edu}

\begin{abstract}
We present a multi-wavelength study of four high redshift blazars, S5 0014+81 ($z=3.37$), CGRaBS J0225+1846 ($z=2.69$), BZQ J1430+4205 ($z=4.72$), and 3FGL J1656.2$-$3303 ($z=2.40$), using the quasi-simultaneous data from \swift, \nustar, and \fermi-Large Area Telescope (LAT) and also the archival {\it XMM-Newton} observations. Other than 3FGL J1656.2$-$3303, none of the sources were known as \gm-ray emitters and our analysis of $\sim$7.5 years of LAT data reveals the first time detection of the statistically significant \gm-ray emission from CGRaBS J0225+1846. We generate the broadband spectral energy distributions (SED) of all the objects, centering at the epoch of \nustar~observations and reproduce them using a one zone leptonic emission model. The optical$-$UV emission in all the objects can be explained by the radiation from the accretion disk, whereas, X-ray to \gm-ray window of the SEDs are found to be dominated by the inverse Compton scattering off the broad line region photons. All of them host billion solar mass black holes. Comparing the accretion disk luminosity and the jet power of these sources with a large sample of blazars, we find them to occupy high disk luminosity-jet power regime. We also investigate the X-ray spectral properties of the sources in detail with a major focus on studying the causes of soft X-ray deficit, a feature generally seen in high redshift radio-loud quasars. We summarize that this feature could be explained based on the intrinsic curvature in the jet emission rather than due to external effects predicted in the earlier studies, such as host galaxy and/or warm absorption.
\end{abstract}


\section{Introduction}\label{sec:intro}
Blazars constitutes a special class of radio-loud active galactic nuclei (AGN) in which the relativistic plasma shoots out from the central core of the galaxy, in the form of jets, towards the line of sight to the observer. The flux enhancement due to special relativity and their peculiar orientation (or so called relativistic beaming) makes blazars visible even at high redshifts ($z>$2) and thus they can be used to explore the young Universe. Blazars are classified as flat spectrum radio quasars (FSRQs) and BL Lac objects based on the rest-frame equivalent width (EW) of their broad optical emission lines, with FSRQs having EW$>$5\AA. Among blazars, FSRQs are known upto $z>$5 \citep[e.g.,][]{2004ApJ...610L...9R}.

 The broadband spectral energy distribution (SED) of blazars exhibits a typical double hump structure that is generally explained in terms of synchrotron (covering from radio to UV$-$X-rays) and inverse Compton (IC, contributing from X-rays to \gm-rays) emission from a single population of energetic electrons present in the jet \citep[e.g.,][]{1998MNRAS.301..451G}. It has been found that with increasing luminosity the SED peaks shift to lower frequencies \citep[][]{1998MNRAS.299..433F}. The shift of the synchrotron peak to lower energies leaves the accretion disk radiation (the big blue bump) naked and is observed in the optical-UV SEDs of many FSRQs \citep[see, e.g.,][]{2013MNRAS.433.2182S}. On the other hand, the shift of the high energy peak from $\sim$GeV to $\sim$MeV band makes the high redshift blazars brighter at hard X-rays than in the \gm-ray band and this suggests that hard X-ray telescopes are better instruments to constrain the non-thermal jet properties of high redshift blazars \citep[][]{2010MNRAS.405..387G}. Thanks to the excellent sensitivity of recently launched hard X-ray focusing telescope {\it Nuclear Spectroscopic Telescope Array} \citep[\nustar;][]{2013ApJ...770..103H}, it is now possible to explore the rising part of the Compton component of high redshift FSRQs in detail. Furthermore, the recently released Pass 8 data from \fermi-Large Area Telescope (\fermi-LAT) has an increased sensitivity at low energies \citep[][]{2013arXiv1303.3514A} which is important for high redshift objects. Though weak in the \gm-ray window of the SED, a significant \gm-ray detection from high redshift FSRQs will further constrain the location of the IC peak and thus a more accurate measurement of their physical properties. 
 
In many high redshift quasars (and also in blazars) a spectral flattening at soft X-ray energies have been noticed \citep[$E\lesssim2-4$ keV, see, e.g.,][and references therein]{2013ApJ...774...29E}. Earlier, it was attributed to the presence of absorbing material in the source environment, however, the low level of optical-UV extinction observed in a number of sources is inconsistent with this interpretation. A hypothesis of the presence of a `warm absorber' was also proposed to explain the observed soft X-ray dip \citep[e.g.,][]{2004MNRAS.350L..67W,2004MNRAS.350..207W}. Based on \swift/GRB data, \citet[][]{2011ApJ...734...26B} proposed a diffuse intergalactic medium dominated absorption scenario. However, this feature is more frequently observed in radio-loud quasars than radio-quiet objects \citep[e.g.,][]{2013ApJ...774...29E} and thus requires a connection between X-ray absorption and jet linked activities. More recently, it has been argued that the observed X-ray flattening could be due to low-energy roll-off of the external Compton (EC) process \citep[e.g.,][]{2007MNRAS.382L..82G,2007ApJ...665..980T,2009AdSpR..43.1036F}.
 
 Here we present the results of a multi-wavelength study of the four high redshift FSRQs, namely, S5 0014+81 (hereafter J0014+81; $z=3.37$), CGRaBS J0225+1846 (hereafter J0225+1846; $z=2.69$), BZQ J1430+4205 (hereafter J1430+4205; $z=4.72$), and 3FGL J1656.2$-$3303 (hereafter J1656$-$3303; $z=2.40$). Our primary goal is to study the broadband physical properties of these high redshift blazars with a major focus on their X-ray characteristics. All of them have been observed from \nustar~along with the simultaneous monitoring from \swift~\citep{2004ApJ...611.1005G}, thus providing a unique opportunity to study their 0.3$-$79 keV X-ray properties. It should be noted that the objects J0014+81 and J0225+1846 have recently been studied by \citet{2015arXiv151008849S} using the same data set. Our findings are similar to them but we also include the recent {\it XMM-Newton} and \fermi-LAT observations and place more emphasis on the X-ray spectral properties. In Section \ref{sec:basic}, we briefly report the basic information about all the four blazars. Section \ref{sec:data_red} is devoted to the adopted data reduction procedures and the obtained results are presented in Section \ref{sec:results}. We discuss our findings in Section \ref{sec:dscsn} and conclude in Section \ref{sec:summary}. We use a $\Lambda$CDM cosmology with the Hubble constant $H_0=71$~km~s$^{-1}$~Mpc$^{-1}$, $\Omega_m = 0.27$, and $\Omega_\Lambda = 0.73$.

\section{General Physical Properties}\label{sec:basic}
\citet[][]{1981AJ.....86..854K} first reported the discovery of J0014+81 and subsequently noted it as an optically luminous quasar \citep[][]{1983ApJ...275L..33K}. Very long baseline interferometry observations reveals the marginal detection of superluminal patterns \citep[][]{2012ApJ...758...84P} and also its optical radiation is not polarized \citep[][]{2011ApJS..194...19W}. \citet[][]{2009MNRAS.399L..24G} derived its central black hole mass to be as high as $\sim$4 $\times$ 10$^{10}$ $M_{\odot}$, however, later it was found to be significantly overestimated \citep[][]{2015arXiv151008849S}. J0014+81 is included in 70 months \swift-Burst Alert Telescope (BAT) hard X-ray catalog \citep{2013ApJS..207...19B} and was observed by \nustar~on 2014 December 18 and 2015 January 23.

J0225+1846 was discovered as a flat spectrum radio bright object by \citet[$F_{\rm 5~GHz}=595$ mJy;][]{1983ApJS...51...67L}. Subsequently, it was identified as a high redshift blazar in the {\it ROSAT} Bright Survey \citep{2000AN....321....1S}. Its is a hard X-ray bright source and is included in the 70 months \swift-BAT catalog. It was observed by \nustar~on 2014 December 24 and 2015 January 18 along with the simultaneous monitoring from \swift. Moreover, two $\sim$100 ksec observations were done using {\it XMM-Newton} \citep[][]{2001A&A...365L...1J} in 2013 January. This source is not present in any \gm-ray catalogs but was postulated as a candidate \gm-ray emitting blazar by \citet{2008ApJS..175...97H}.

At a redshift of 4.72, J1430+4205 is a radio-loud \citep[$F_{\rm 5~GHz}=230$ mJy;][]{2007ApJ...658..203H} and an extremely X-ray luminous blazar \citep[e.g.,][]{1997MNRAS.291L...5F,1998MNRAS.295L..25F,1998MNRAS.294L...7H}. \citet{1999MNRAS.308L...6F} reported the detection of a significant radio and X-ray flux variability from this source and postulated the presence of $\sim$billion solar mass black hole at the center. Later, {\it XMM-Newton} observations confirmed the presence of soft X-ray flattening in the X-ray spectrum of this object \citep[][]{2004MNRAS.350L..67W}, a feature typically observed in many high redshift radio-loud quasars \citep[e.g.,][]{2005MNRAS.364..195P}. J1430+4205 was monitored by \nustar~on 2014 July 14, thereby making it the second most distant source observed by \nustar~after B2 1023+25 \citep[$z=5.28$;][]{2013ApJ...777..147S}.

The source J1656$-$3303 was discovered with the \swift~BAT during its first year of survey operation \citep{2006ATel..799....1O}. This object lies close to the galactic plane (galactic latitude $\sim$6$^{\circ}$.3). Using the \swift~X-ray Telescope \citep[XRT;][]{2005SSRv..120..165B} observations, \citet{2006ATel..835....1T} suggested it to be an extragalactic source. Later, a detailed multi-wavelength study of this source was done by \citet{2008A&A...480..715M} with a major focus on its optical spectroscopic properties. They confirmed the extragalactic nature of J1656$-$3303 and concluded it to be a high redshift blazar at $z=2.40$. It is included in the recently released third catalog of \fermi-LAT detected sources \citep[3FGL;][]{2015ApJS..218...23A}. It did not appear in the clean sample of the third catalog of \fermi-LAT detected AGNs  \citep[3LAC;][]{2015ApJ...810...14A} because of its close proximity to the galactic plane. This source was simultaneously observed by \nustar~and \swift~on 2015 September 27.

\section{Multiwavelength observations and Data Reduction}\label{sec:data_red}
\subsection{{\it Fermi}-Large Area Telescope Observations}\label{subsec:fermi}
Three out of four sources studied here, J0014+81, J0225+1846, and J1430+4205, are not present in any \gm-ray catalogs. Therefore, by utilizing the availability of recently released Pass 8 data from \fermi-LAT, we not only study the \gm-ray properties of the known \gm-ray emitter J1656$-$3303 but also search for the possible detection of remaining three sources in the LAT energy range. We consider the first 89 months of \fermi-LAT data (MJD 54683$-$57391 or 2008 August 4 to 2016 January 4) and follow the standard procedure as described in the online documentation\footnote{http://fermi.gsfc.nasa.gov/ssc/data/analysis/documentation/}. In the energy range of 0.1$-$300 GeV, we select only events belonging to SOURCE class (corresponding to {\tt evclass=128}) and use a relational filter ``\texttt{DATA$\_$QUAL$>$0}'', \&\& ``\texttt{LAT$\_$CONFIG==1}'' to select good time intervals. To avoid the contamination from Earth limb \gm-rays, we reject all the events with zenith angle larger than 90$^{\circ}$. We define the region of interest (ROI) as a circle of 15$^{\circ}$ radius centered at the target source to perform the likelihood fitting. All the objects lying within the ROI and present in 3FGL catalog are considered and the spectral parameters of those lying within 10$^{\circ}$ are left free to vary during the fitting. The parameters of remaining sources are fixed to the 3FGL catalog values. We use a binned likelihood method and compute the significance of the $\gamma$-ray signal by means of the maximum likelihood (ML) test statistic TS =  2$\Delta \log (\mathcal{L}$) where $\mathcal{L}$ represents the likelihood function, between models with and without a point source at the position of the source \citep[][]{1996ApJ...461..396M}. To search for the presence of additional faint objects that could be present in the data but not in 3FGL catalog, we generate the residual TS maps, after performing the first round of likelihood fitting. If present, these newly detected objects are then modeled with a power law model and another round of fitting is performed. We generate another residual TS map to ascertain that no sources are left to model. We then removed all the sources having TS$<$25 to perform further temporal and spectral studies. To generate both light curve and spectrum, the source is considered to be detected if TS$>$9. For bins with TS$<$9 and or $\Delta F_{\gamma}/F_{\gamma} > 0.5$, where $\Delta F_{\gamma}$ is the error estimate in the flux $F_{\gamma}$, we calculate 2$\sigma$ upper limit. In this work, all the errors associated with the LAT data analysis are 1$\sigma$ statistical uncertainties, unless specified.

\subsection{\nustar~Observations}\label{subsec:nustar}
The blazar J0014+81 was monitored two times by \nustar~each on 2014 December 21 (obs ID 60001098002) and 2015 January 23 (obs ID 60001098004) for a net exposure of 36.4 ksec and 31 ksec, respectively. J0225+1846 was also observed twice, each on 2014 December 24 (obs ID 60001101002) and 2015 January 18 (obs ID 60001101004) for net exposure times of 32.0 and 37.5 ksec, respectively. J1430+4205 was monitored on 2014 July 14 (obs ID 60001103002) for 49.2 ksec, whereas, J1656$-$3303 was observed by \nustar~on 2015 September 27 (obs ID 60160657002) for a net exposure time of 21.1 ksec.

The downloaded data sets for Focal Plane Module A (FPMA) and Focal Plane Module B (FPMB) are analyzed using the \nustar~Data Analysis Software (NUSTARDAS) version 1.5.1. The task {\tt nupipeline} is used to clean and calibrate the events data files using standard filtering criteria and \nustar~CALDB updated on 2015 October 8. Source and background spectra, along with the ancillary and response matrix files, are generated using the task {\tt nuproducts}. To generate the source spectrum, we select the source region as a circle of 30$^{\prime\prime}$ radius centered at the target source. Background spectrum is extracted from a circular region of 70$^{\prime\prime}$ on the same chip, free from source contamination. We bin the source spectrum to have a signal-to-noise  ratio of 6, and oversample the spectra by a factor of 3.

\subsection{{\it XMM-Newton} Observations}\label{subsec:xmm}

J0014+81 was observed with \xmm\ on 2001 August 23, for a net exposure of 42.9~ksec (obs ID 0112620201). J0225+1846 was observed three times, once in 2003 (July 25, obs ID 0150180101) and twice in 2013 (January 13 and 15, obs IDs 0690900101 and 0690900201), for exposures of 22.2, 108.0, and 96.7~ksec, respectively. J1430+4205 was also observed three times, on 2002 December 09 (obs ID 0111260101), 2003 January 17 (obs ID 0111260701), and 2005 June 05 (obs ID 0212480701) for exposures of 18.9, 14.6 and 19.7~ksec. Finally, J1656-3303 was observed on 2009 September 11 (obs ID 0601741401) for net exposure of 22.6~ksec.

We use the \xmm\ Science Analysis Software (SAS) version 15.0.0 to reduce the data. We use the {\tt rgsproc} task to reduce the RGS spectra and the {\tt epproc} task to produce the EPIC-pn event files, which we then filter for high background using {\tt evselect}. We extract source spectra from 30$^{\prime\prime}$ circular regions, and background spectra from 60$^{\prime\prime}$ circular regions on the same chip, avoiding contaminating sources. The spectra are binned using {\tt specgroup} to a signal-to-noise ratio of 6, after background subtraction, and to oversample the data by a factor of 3.

\subsection{{\it Swift} Observations}\label{subsec:swift}
\swift~observations were carried out in conjunction with \nustar~monitoring for all four blazars. It observed J0014+81 on 2014 December 21 (obs ID 00080003001, net exposure 6.5 ksec) and 2015 January 23 (obs ID 00080003002, net exposure 6.6 ksec). The source J0225+1846 was monitored on 2014 December 24 (obs ID 00080243001, net exposure 4.9 ksec) and 2015 January 18 (obs ID 00080243002, net exposure 5.1 ksec). \swift~observed J1430+4205 on 2014 July 13 (obs ID 00080752002, net exposure 7.5 ksec) and J1656$-$3303 on 2015 September 27 (obs ID 00081202001, net exposure 6.8 ksec).

\swift-XRT observations of all the sources were carried out using the most sensitive photon counting mode (standard grade selection 0$-$12). The event files are calibrated and cleaned with the task {\tt xrtpipeline} and using latest calibration files. Calibrated and cleaned event files are summed using {\tt xselect} and the resultant summed event files are used to extract energy spectrum. We select the source region as a circle of 55$^{\prime\prime}$ radius, whereas, events for the background spectra are extracted from an annular ring with inner and outer radii of 110$^{\prime\prime}$ and 210$^{\prime\prime}$, respectively, both centered at the position of the source of interest. We combine the exposure maps using the tool {\tt ximage} which takes into account for CCD defects and PSF losses and generate the ancillary response files using {\tt xrtmkarf}. The source spectra are binned to have a signal-to-noise ratio of 6.

The \swift~UltraViolet Optical Telescope \citep[UVOT;][]{2005SSRv..120...95R} has observed the sources of interest in all the six filters ($V$, $B$, $U$, $W$1, $M$2, and $W$2). To improve the signal to noise ratio, we add the UVOT snapshots using the tool {\tt uvotimsum}. The source counts are extracted from a circular region of 5$^{\prime\prime}$ centered at the source of interest, while background is selected as a circle of 30$^{\prime\prime}$ radius from a nearby source-free region. The task {\tt uvotsource} is used to extract the magnitudes which are then corrected for Galactic reddening \citep{2011ApJ...737..103S}. The de-absorbed magnitudes are converted to flux units using the zero point and calibrations of \citet{2011AIPC.1358..373B}.

\section{Results}\label{sec:results}
\subsection{Gamma-ray Properties}\label{subsec:fermi}
\begin{figure*}
\hbox{
      \includegraphics[width=9cm]{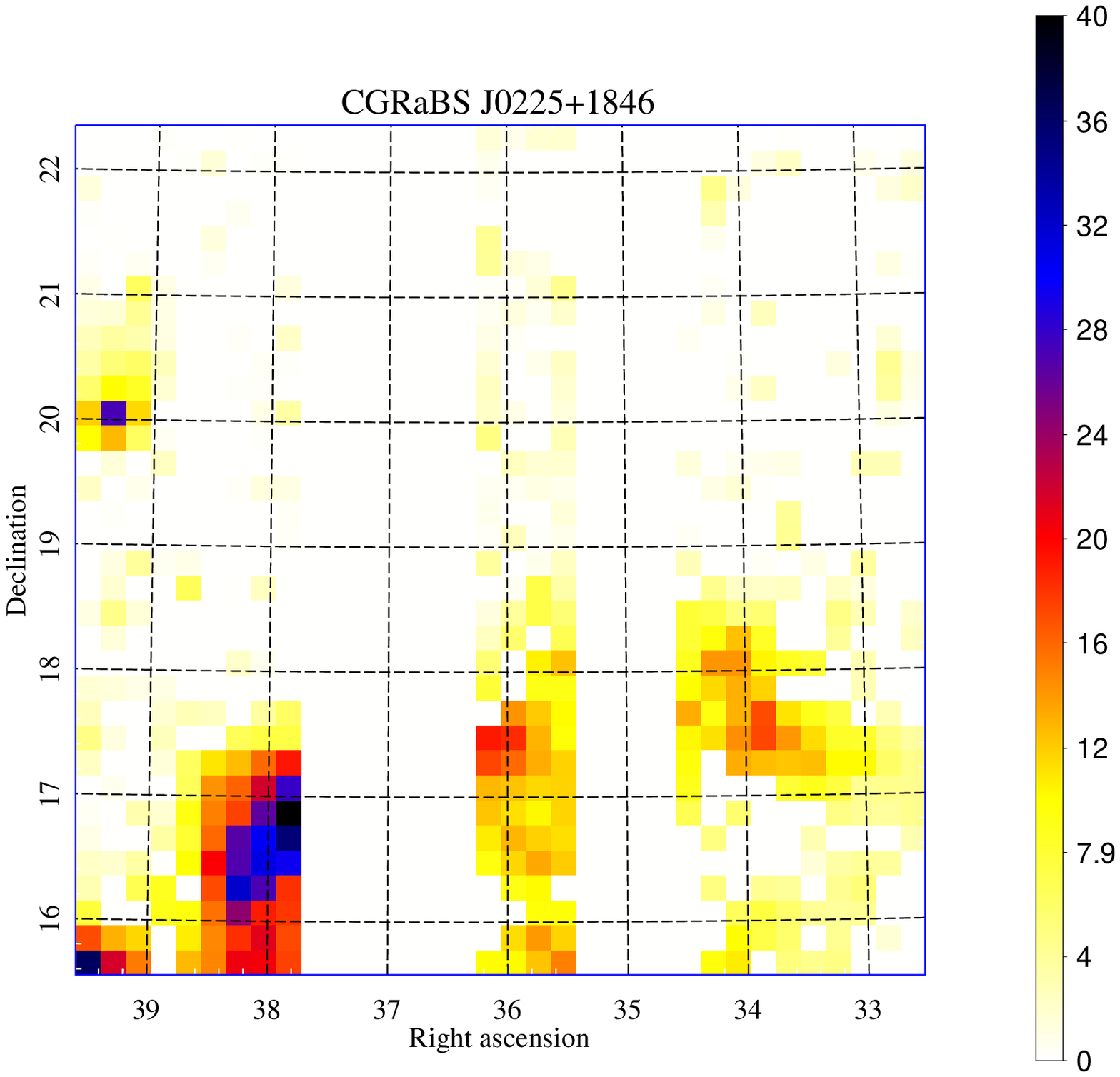}
      \includegraphics[width=9cm]{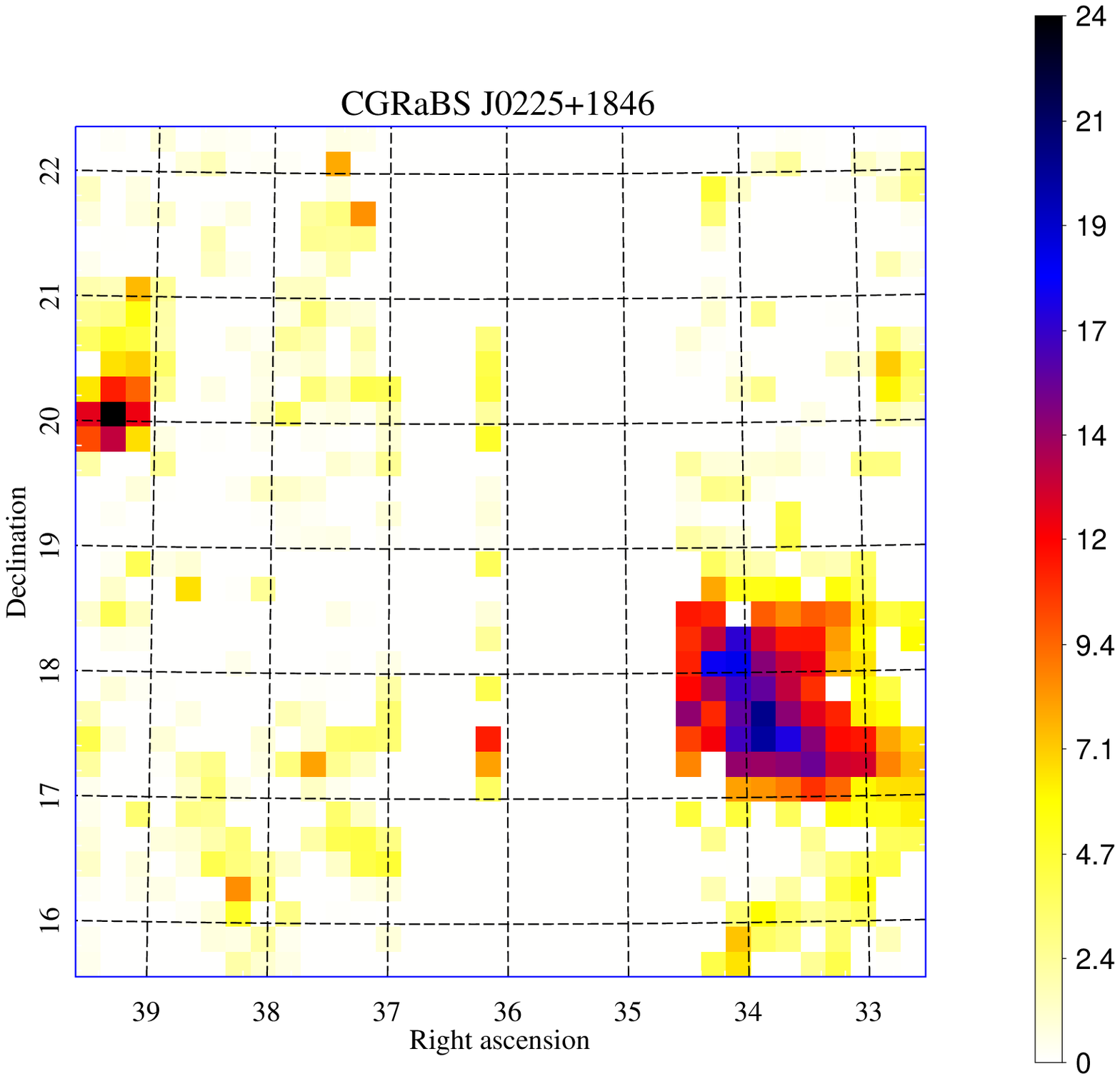}
     }
 \hbox{\hspace*{4cm}
    \includegraphics[width=9cm]{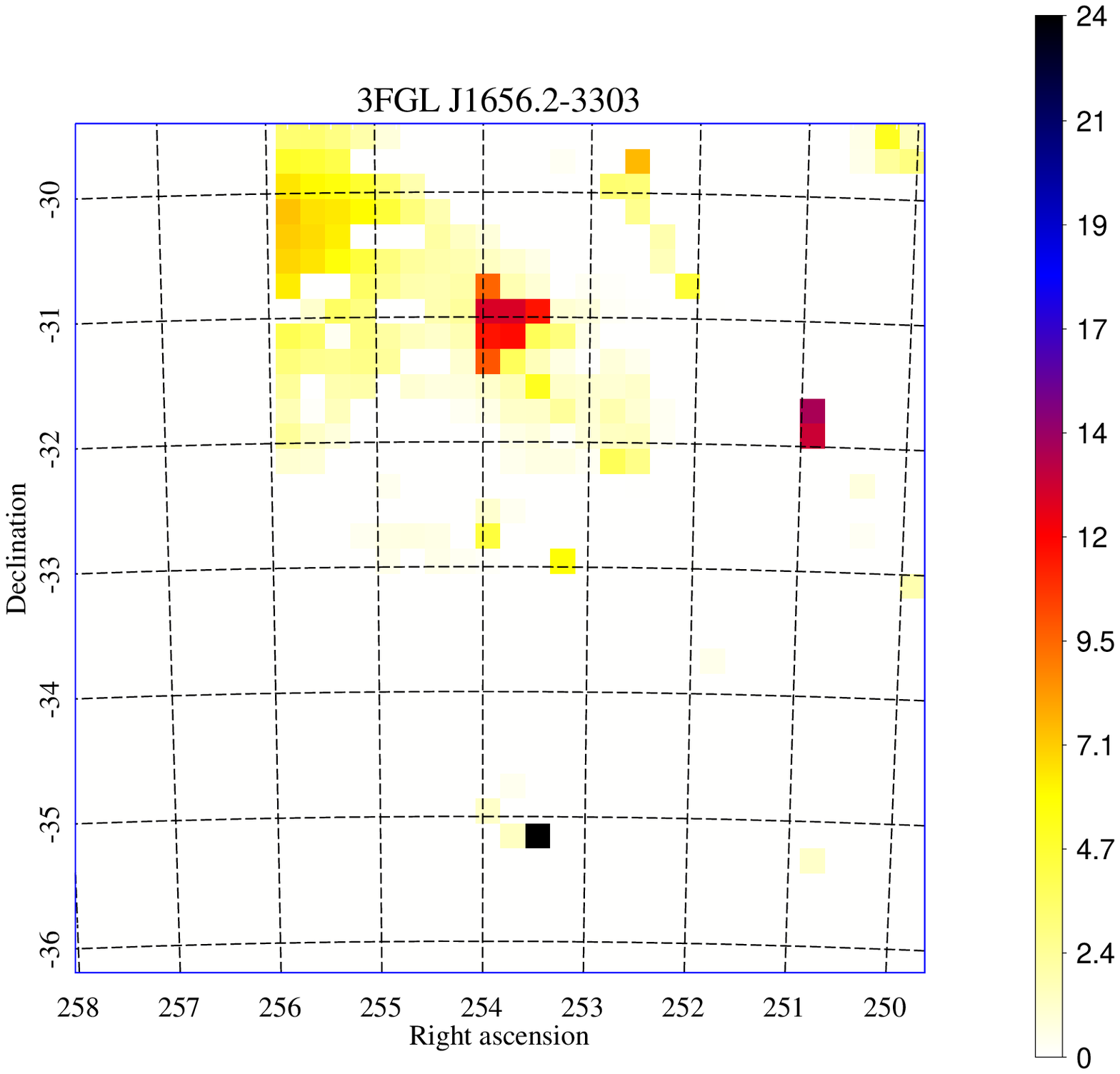}
    }
\caption{Residual TS maps of 0.1$-$300 GeV events centered at J0225+1846 (top) and J1656$-$3303 (bottom), the sources from which a significant \gm-ray emission is detected. The TS map of J0225+1846 in top left panel indicates the presence of unmodeled objects with TS$>$25, whereas, top right panel corresponds to the same field but after considering unmodeled objects. On the other hand, no unmodeled source with significant \gm-ray emission is detected in the TS map of J1656$-$3303.}\label{fig:tsmap}
\end{figure*}

None of the blazars under consideration, except J1656$-$3303, are present in any \gm-ray catalogs. Therefore, using the good quality Pass 8 LAT data, which is more sensitive at low energies, we search for possible \gm-ray signal from all the four sources. This is done by performing an average analysis of $\sim$7.5 years of LAT data. Other than J1656$-$3303, we find a statistically significant \gm-ray emission from J0225+1846 (TS$\approx$188, $\sim$13$\sigma$ detection). To ensure that the observed \gm-ray emission is associated with target blazars and not with any other unmodeled objects lying close to the sources of interest, we generate their residual TS maps and show them in Figure \ref{fig:tsmap}. As can be seen, the residual TS map of J0225+1846 suggests for the presence of few unmodeled sources (with TS$>$25) and we properly consider them in the analysis. The residual TS map of the same field, after taking unmodeled objects into account, verifies that no other sources (with TS$>$25) are left to model. The \gm-ray spatial location of J0225+1846 is optimized using the tool {\tt gtfindsrc} and derived as R.A., decl. = 36$^{\circ}$.177, 18$^{\circ}$.859 (J2000) with a 95\% error circle radius of 0$^{\circ}$.21 degrees. We cross check in NRAO VLA Sky Survey \citep[][]{1998AJ....115.1693C} for the presence of additional radio sources within the 95\% contour and found a total of 10 sources with J0225+1846 being the brightest ($F_{\rm 1.4 GHz}=460.8$ mJy). Then, we search in the NASA/IPAC Extragalactic Database\footnote{http://ned.ipac.caltech.edu/} for multi-wavelength counterparts of these radio sources and find no information for any objects other than J0225+1846. This indicates the close association of J0225+1846 with the \gm-ray emitter. Furthermore, the TS map of J1656$-$3303 does not reveal any unmodeled sources and therefore the model used during the analysis is an accurate representation of the observed \gm-rays, with no new \gm-ray point source being present. Apart from these, LAT data analysis of J0014+81 and J1430+4205 resulted in TS = 1.47 and 5.62, respectively. This suggests the non-detection of J0014+81 and only a marginal $\sim$2$\sigma$ detection of J1430+4205 in the \gm-ray band. Two models, a power law ($N(E) = N_0(E/E_0)^{-\Gamma_{\gamma}}$, where $\Gamma_{\gamma}$ is the photon index) and a logParabola ($N(E) = N_0(E/E_{\rm pivot})^{-\alpha-\beta~{\rm log}(E/E_{\rm pivot})}$, where $\alpha$ is the photon index at $E_{\rm pivot}$, $\beta$ is the curvature index, and $E_{\rm pivot}$ is pivot energy fixed at 300 MeV) are applied to the \gm-ray spectra of J0225+1846 and J1656$-$3303. The results of the average LAT data analysis are presented in Table \ref{tab:avg_gamma}.

\begin{figure*}
      \includegraphics[width=\columnwidth]{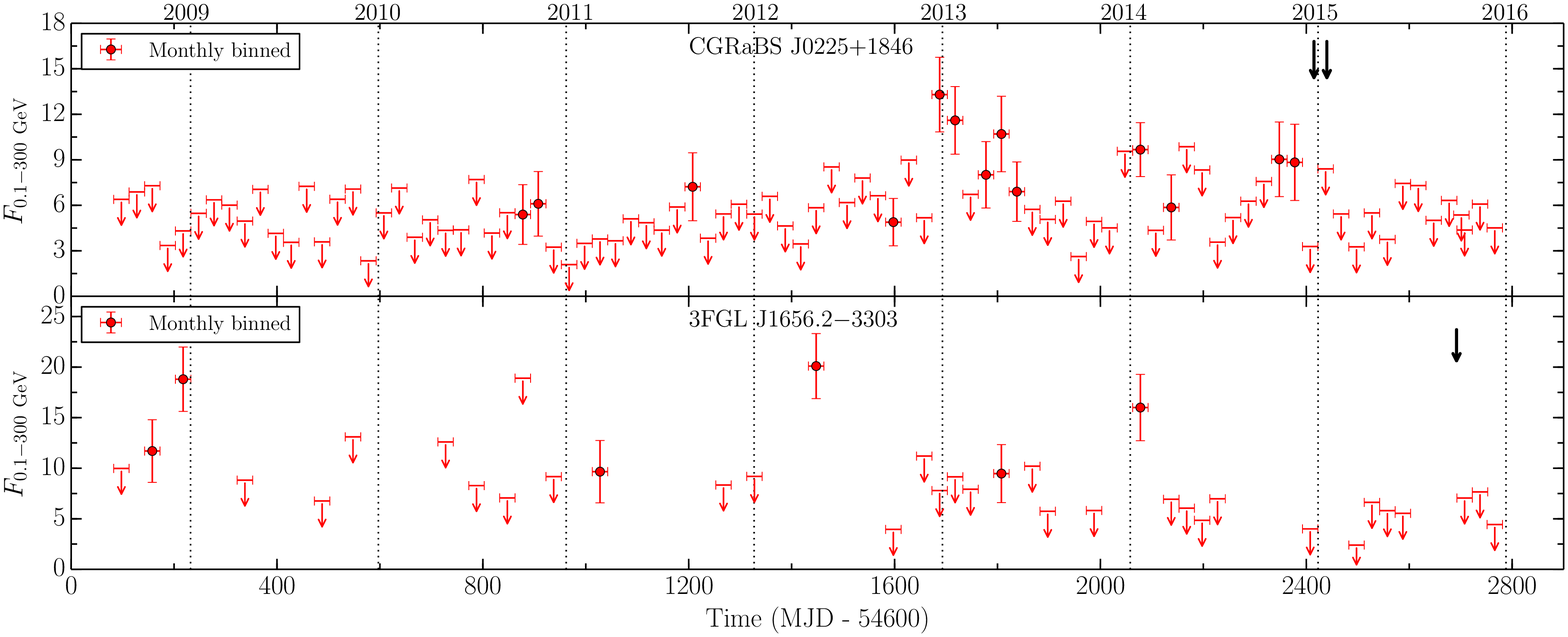}
\caption{Temporal behavior of J0225+1846 (top) and J1656$-$3303 (bottom) in 0.1$-$300 GeV energy range, as observed by \fermi-LAT. The fluxes are in units of 10$^{-8}$ \phflux. Red downward arrows represent 2$\sigma$ flux upper limits, whereas, black downward arrows denote the time of \nustar~monitoring. Note that the time bins with TS$<$1 are not shown. Vertical dotted lines correspond to the beginning of the calender years.}\label{fig:fermi_lc}
\end{figure*}

We also study the \gm-ray temporal behavior of J0225+1846 and J1656$-$3303. This is done by generating monthly binned \gm-ray light curves which we show in Figure \ref{fig:fermi_lc}. Both of them are faint in \gm-rays and only a moderate brightening of J0225+1846 is seen around the beginning of 2013. On the other hand, J1656$-$3303 is occasionally detected by LAT without any flaring activity.

\subsection{X-ray Properties}\label{subsec:x-ray}

We fit the \nustar~(3$-$79 keV), \swift-XRT (0.3$-$10 keV), and \xmm~(0.3$-$10 keV for EPIC-pn and 0.5$-$2 keV for RGS) spectra using \textsc{xspec} \citep{Arnaud96} version 12.9.0j.
In all cases we include Galactic absorption, taken from the online tool of \citet{Willingale13}, of $2.19\times 10^{21}$~cm$^{-2}$ for J0014+81, $1.68\times 10^{21}$~cm$^{-2}$ for J0225+1486, $1.20\times 10^{20}$~cm$^{-2}$ for J1430+4205, and $3.35\times 10^{21}$~cm$^{-2}$ for J1656-3303.
We use \texttt{wilm} abundances \citep{Wilms00} and \texttt{vern} cross sections \citep{Verner96}.

\subsubsection{Broad-band 0.3$-$79 keV spectra}
The most useful spectra for determining the nature of the change in slope in the X-ray band are the simultaneous \swift -XRT and \nustar\ spectra, which cover the energy range from $\sim0.3$--79~keV. We fit the same three models to each spectrum: a powerlaw absorbed by Galactic absorption only (M1, \emph{tbabs * powerlaw} in \textsc{xspec}); a powerlaw absorbed by Galactic absorption and absorption intrinsic to the source (M2, \emph{tbabs * ztbabs * powerlaw}); and a broken power-law, absorbed by Galactic absorption only (M3, \emph{tbabs * bknpower}). These correspond to a spectrum with no curvature, a spectrum where the curvature is due to absorption of the source by the host galaxy, and a spectrum where the curvature is intrinsic. Two of the sources (J0014+81 and J0225+1846) were observed twice with \nustar\ and \swift . In the case of J0014+81 the two spectra are indistinguishable, so we fit them simultaneously with the same model. For J0225+1846 the observations are further apart and show significant spectral evolution. We, therefore, fit them simultaneously but allow the model parameters to vary between the two observations.

\begin{figure*}
\centering
\includegraphics[width=12cm]{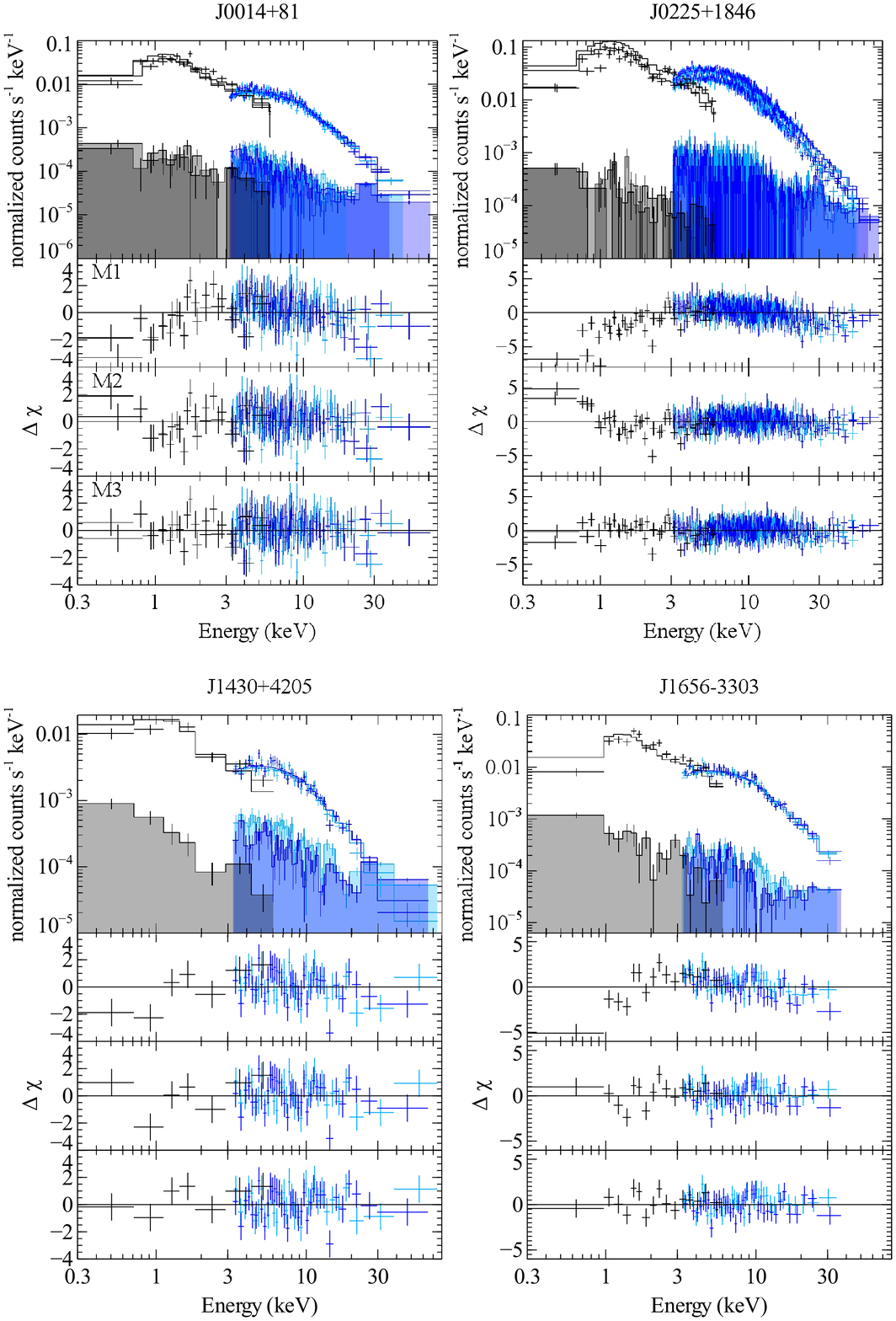}
\caption{$\Delta\chi$ residuals from the fits to the joint \nustar\ (blue) and XRT (black) spectra of the four blazars. In each case, the three models (power-law, absorbed power-law, and broken power-law) are shown in order from top to bottom. The top panel in each plot shows the data, fit with M1 (power-law), and the shaded region shows the background spectrum.}
\label{fig_nustar_xrt_residuals}
\end{figure*}

The best fit parameters for each model and object are given in Table~\ref{table_xrtnustar} and the fit residuals are shown in Figure~\ref{fig_nustar_xrt_residuals}. In each case, the quality of the fit improves from M1--M3, with the absorbed power-law preferred over the unabsorbed power-law, and the broken power-law better still. The improvement in $\chi^2$ from switching from M2 to M3 is marginal for J0014+81, J1430+4205 and J1656-3303, so we check the significance using an F-test. We find chance probabilities of 0.0003, 0.011 and 0.008, respectively (in all cases, the improvement obtained by switching from M1 to M2 or M3 is significant at $\gg3\sigma$). 

We also consider the effect of ionized absorption (warm absorber) models. We replace the neutral absorber in M2 with an ionized absorber (modelled using an \textsc{xstar} grid), and re-fit the data for each source. This improves the fit significantly relative to the neutral absorber only in J0225+1846 ($\chi^2_\nu=1.04$), with a column density of $N_\textrm{H}=5.5_{-0.9}^{+1.7}\times10^{23}$, and ionization of $\log(\xi)=2.8\pm0.1$ (the spectral variability between the two observations is mostly in the power-law).

\subsubsection{XMM-Newton}

We next fit the \xmm\ data for each object. We fit the EPIC-pn spectrum for each source, fitting simultaneously but allowing parameters to vary independently when there are multiple spectra. For J0225+1846, there are also high-resolution RGS spectra, which we include in the fit with model parameters tied to those of the corresponding EPIC-pn spectrum. For this fit we use the higher resolution \emph{tbnew} \citep{Wilms00}. 

As these observations are not taken simultaneously with a hard X-ray instrument the energy band is significantly narrower, making it harder to detect spectral curvature. As such, additional curvature is not significantly detected in one source, J0014+81. This spectrum is well fit with a simple power-law model (M1, $\chi^2/\mathrm{d.o.f.}=112/116$) and the fit is not significantly improved by either additional absorption or a spectral break.

\begin{figure*}
\centering
\includegraphics[width=15cm]{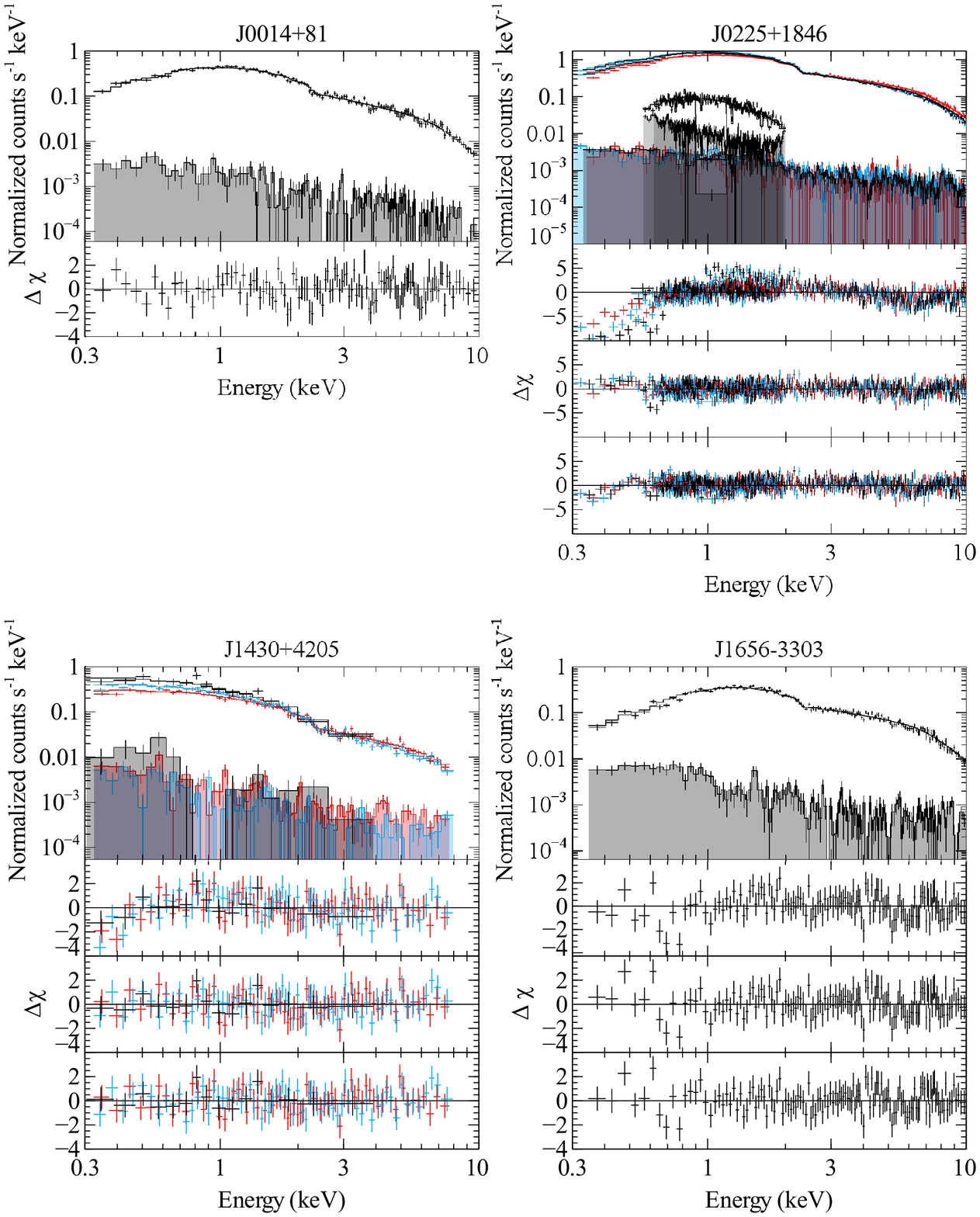}
\caption{$\Delta\chi$ residuals from the fits to the \xmm\ spectra of the four blazars. In each case, the three models (power-law, absorbed power-law, and broken power-law) are shown in order from top to bottom. The top panel in each plot shows the data, fit with M1 (power-law), and the shaded region shows the background spectrum.}
\label{fig_xmm_residuals}
\end{figure*}

The other three sources show evidence of spectral curvature, with a significant improvement in the fit from M1 to M2 or M3. There is only a significant improvement from M2 to M3 in the case of J1656$-$3303 (p=0.004). For J0225+1486 the improvement is marginal (p=0.1) and for J1430+4205 the fit is slightly worse with M3 than M2, although not significantly so. The data and residuals to each model are shown in Figure~\ref{fig_xmm_residuals} and the parameters are given in Table~\ref{table_xmm}.

\subsection{Spectral Energy Distributions}\label{subsec:sed}
\subsubsection{Model Setup}\label{subsubsec:model}

We use a simple one zone synchrotron and inverse Compton emission model fully described in \citet[][]{2009MNRAS.397..985G} and here we discuss it in brief. We assume the emission region is a sphere of radius $R_{\rm blob}$, located at a distance $R_{\rm diss}$ from the central black hole, and moving with a bulk Lorentz factor $\Gamma$. The emitting region is filled with relativistic electrons having a smooth broken power law energy distribution of the following type
\begin{equation}
 Q(\gamma)  \, = \, Q_0\, { (\gamma_{\rm b})^{-n1} \over
(\gamma/\gamma_{\rm b})^{n1} + (\gamma/\gamma_{\rm b})^{n2}}.
\end{equation}
The size of the emission region is constrained by assuming it to cover entire jet cross-section with semi opening angle of the jet as 0.1 rad. The magnetic field is considered as tangled and uniform in the emission region. We consider several sources of thermal radiation externally to the jet: (i) the direct radiation from the accretion disk; (ii) the X-ray corona sandwiching the accretion disk, reprocessing a fraction $f_{\rm cor}$ of the accretion disk luminosity ($L_{\rm disk}$), and having a cut-off power law spectrum ($L_{\rm cor}(\nu)\propto \nu^{-\alpha_{\rm cor}}\exp(-\nu/\nu_{\rm c})$); (iii) the broad line region (BLR), assumed to reprocess a fraction $f_{\rm BLR}$ of $L_{\rm disk}$ from a spherical shell located at a distance $R_{\rm BLR} = 10^{17} L^{1/2}_{\rm disk,45}$ cm, where $L_{\rm disk,45}$ is the disk luminosity in units of 10$^{45}$ erg s$^{-1}$; (iv) and the IR emission from a dusty torus located at a distance $R_{\rm torus} = 10^{18} L^{1/2}_{\rm disk,45}$ cm and re-emit a fraction $f_{\rm torus}$ of the accretion luminosity. The spectra of both the BLR and the torus are assumed to be a blackbody peaking at the rest-frame frequency of Lyman-$\alpha$ line \citep[e.g.,][]{2008MNRAS.386..945T} and $T_{\rm torus}$, respectively, where $T_{\rm torus}$ is the characteristic temperature of the torus. In the comoving frame, we appropriately evaluate the radiative energy densities of these components and used them to derive the EC spectrum, and then transform into observer frame. We estimate the power that the jet carries in the form of magnetic field ($P_{\rm mag}$), radiation ($P_{\rm rad}$), relativistic electrons ($P_{\rm ele}$), and cold protons ($P_{\rm kin}$). The last quantity, i.e., the kinetic jet power, is derived by assuming equal number density of protons and relativistic electrons \citep[e.g.,][]{2008MNRAS.385..283C}.

The flux produced by the accretion disk is estimated by assuming a standard \citet{1973A&A....24..337S} disk having a multi-temperature radial profile as follows \citep[][]{2002apa..book.....F}
\begin{equation}
T \, =\, {  3 R_{\rm Sch}  L_{\rm disk }  \over 16 \pi\eta_{\rm acc}\sigma_{\rm MB} R^3 }  
\left[ 1- \left( {3 R_{\rm Sch} \over  R}\right)^{1/2} \right]^{1/4},
\end{equation}
where $R_{\rm Sch}$ is the Schwarzschild radius and $\eta_{\rm acc}$ is the accretion efficiency. This is used to reproduce the thermal radiation (the big blue bump) observed at optical-UV energies and it constrains both $L_{\rm disk}$ and the central black hole mass of the source. There are only two parameters to be fitted, the accretion rate $\dot{M}_{\rm acc}$ and the black hole mass, once we assume a typical value of the accretion efficiency ($\eta_{\rm acc}=10$\%). The former can be derived from the intrinsic accretion disk luminosity $L_{\rm disk}=\eta_{\rm acc}\dot{M}_{\rm acc}c^2$ with $L_{\rm disk}$ being an observable parameter (provided the peak of the big blue bump is visible in the SED). This leaves only black hole mass as a free parameter and can be calculated from the optical-UV SED fitting.
 
\subsubsection{SED Modeling Results}\label{subsubsec:sed_results}
The broadband SEDs of all the four blazars are generated using quasi-simultaneous data from \swift, \nustar, and \fermi-LAT. None of the sources are detected by LAT even at 3$\sigma$ level in the month of \nustar~observations and therefore we calculate their respective 2$\sigma$ upper limits. The generated SEDs are reproduced by the model presented in Section~\ref{subsubsec:model}. For modeling of the SEDs, we adopt the following assumptions: the spectral shape of the X-ray corona emission is assumed to be flat ($\alpha_{\rm cor}$ = 1) and the cut-off energy is fixed at 150 keV \citep[][]{2009MNRAS.397..985G}. Fractions of the accretion disk luminosity re-emitted by the X-ray corona, the BLR, and the dusty torus are considered as $f_{\rm cor}=0.3$, $f_{\rm BLR}=0.1$, and $f_{\rm torus}=0.5$, respectively. The results of the SED generation and modeling are presented in Figure \ref{fig:sed_fit} and the associated modeling parameters are given in Table \ref{tab:sed_param} and \ref{tab:jet_power}.

\section{Discussion}\label{sec:dscsn}
A study of the high redshift blazars offers a unique opportunity to understand the physical properties of jetted sources at the extreme end of the accretion disk luminosity and the jet power. Since blazars emit a major fraction of their bolometric luminosity in the form of high energy X-ray and \gm-ray radiation, it is important to have a good quality sensitive monitoring at these energies. In this regard, the observations from the facilities like \fermi-LAT, \nustar, and {\it XMM-Newton} are crucial to learn the radiative processes powering the jet of these distant sources.

\subsection{Gamma-ray Emission}\label{subsec:gm_ray_dscsn}
The high redshift blazars are, in general, weak in the \gm-ray band because of the $k$-correction effect (for increasing redshifts) and also due to intrinsic shift of the IC peak at lower frequencies as their bolometric luminosity increases \citep[e.g.,][]{1998MNRAS.301..451G}. Three out of four sources studied here were never detected in the \gm-ray band, thus supporting the above hypothesis. However, thanks to the recently released Pass 8 data from \fermi-LAT, which is more sensitive at lower energies and thus important for high redshift objects, it is now possible to search distant blazars in the \gm-ray band and constrain their SEDs in a far better way than in the past. We search for the detection of significant \gm-ray emission from all the four high redshift blazars studied in this work (including J1656$-$3303 which is included in 3FGL catalog) and report J0225+1846 as a newly detected \gm-ray emitter at $\sim$13$\sigma$ significance. This is the first report of the discovery of the \gm-ray emission from this object and confirms the prediction of \citet{2008ApJS..175...97H}. We also find a marginal $\sim$2$\sigma$ evidence for the detection of J1430+4205 at \gm-ray energies. Moreover, we search for temporal flux variations among J0225+1846 and J1656$-$3303 and find the former to exhibit a low amplitude \gm-ray flare around the beginning of the year 2013 (see Figure \ref{fig:fermi_lc}). This source has also shown hints of spectral curvature in its $\sim$7.5 years average \gm-ray spectrum.

\subsection{Soft X-ray Flattening}\label{subsec:X_ray_dscsn}
A flattening of the soft X-ray spectrum is observed in all the four blazars considered in this study. In fact, this feature is observed in several high redshift quasars \citep[e.g.,][]{2013ApJ...774...29E} and, in earlier studies, it was attributed to the presence of absorbing material intrinsic to the source environment \citep[e.g.,][]{2001MNRAS.324..628F,2001MNRAS.323..373F}. However, a low level of reddening observed at optical-UV energies does not support the above hypothesis. This is because, in some cases, the column densities derived from the X-ray spectral fitting reaches as high as $N_{\rm H}\approx10^{23}$ cm$^{-2}$ \citep[][]{2001MNRAS.324..628F} and accordingly the reddening in the optical-UV band would be large ($A_{\rm V}\sim100$), which is against the observations. An alternative solution, so called `warm absorber', was proposed by invoking an extreme gas-to-dust ratio, probably because of the high ionization state of the objects \citep[e.g.,][]{2001MNRAS.324..628F}. On the other hand, the recent work by \citet[][]{2007ApJ...665..980T} \citep[see also,][]{2007MNRAS.382L..82G,2009AdSpR..43.1036F} suggests the soft X-ray flattening to be originated from the jet emission.

\begin{figure*}
\centering
\includegraphics[width=14cm]{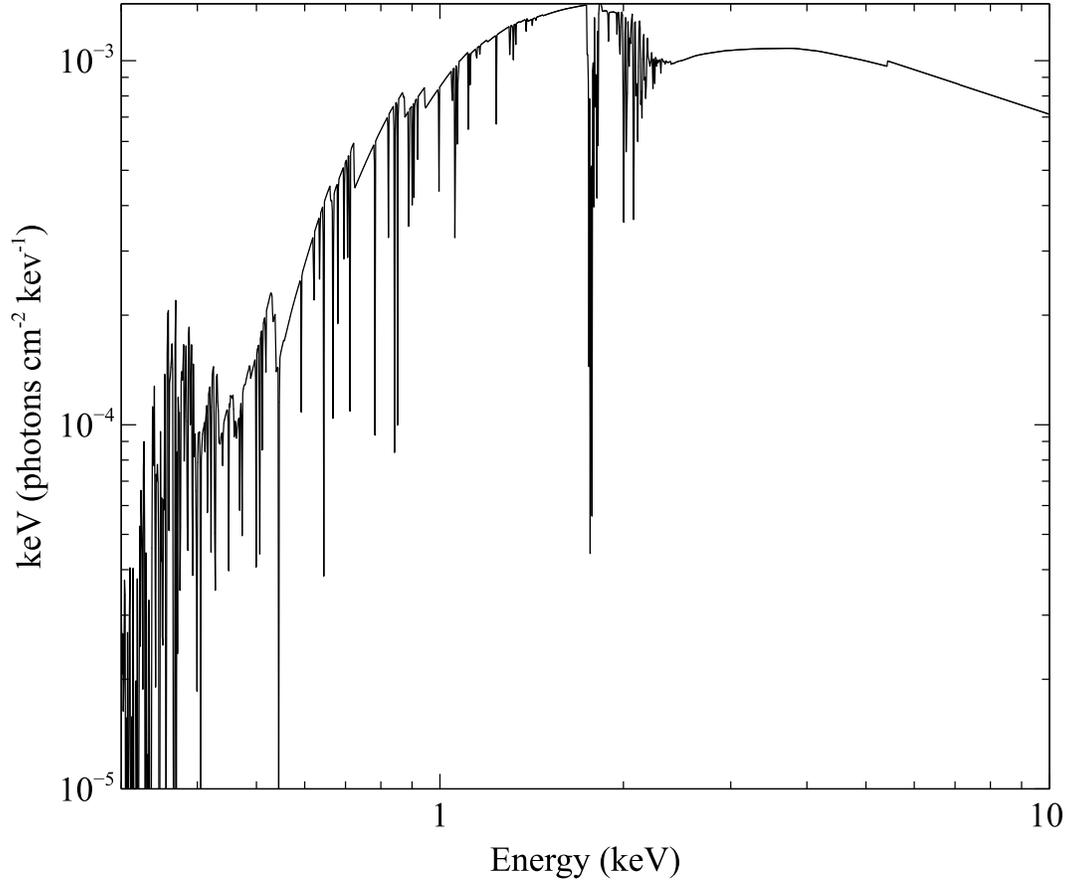}
\caption{Ionized absorption model for the broad-band \nustar /\swift\ spectrum of J0225+1846. Energy scale is in the observer's frame. Strong iron absorption lines are visible, redshifted to around 2~keV, which would be easily detectable in the \xmm\ spectrum.}
\label{fig_0225_model}
\end{figure*}

Switching M2 to use an ionized absorber (modeled with an \textsc{xstar} grid) offers an improved fit to the XRT/\nustar\ spectrum of J0225+1846, although still not as good as the fit with M3 (the broken power-law). Additionally, this model requires an extremely high column density ($\sim5\times10^{23}$ cm$^{-2}$) to account for the observed curvature in this spectrum. The combination of ionization and high column density predicts very strong iron absorption lines around 6--7~keV in the rest frame, redshifted to just below the \nustar\ band (see Figure~\ref{fig_0225_model}). The strength of these features is such that they would be easily detected by the more sensitive EPIC-pn spectrum or the high resolution RGS spectrum in the case of J0225+1846.

More generally, there is a large disagreement between the column densities measured from the broad-band XRT/\nustar\ spectra and the \xmm\ spectra. The column densities from the broad-band spectra are consistently several times larger (over 10 times in the case of J1656$-$3303). If the spectral curvature is caused by absorption, these values should be the same, or at least there should not be a consistent trend if the absorption is variable. However, if the spectral curvature is more gradual and intrinsic to the source, then we would expect exactly this trend, as more curvature is found in the broad-band spectra and thus a higher column is needed to fit the data. A similar trend is observed with the broken power-law model (M3), where the break energy is lower in the \xmm\ spectra and the power-law indices are lower. This is exactly as expected from fitting such a model to a smooth curve.

While it is always possible that the absorbing column to each source has changed between the \xmm\ and \nustar\ observations, it is extremely unlikely that all four sources would change in the same manner. When combined with the better fits to the broad-band spectra found using the broken power-law model, this is very strong evidence that the spectral curvature is not caused by absorption, but is instead an intrinsic part of the blazar X-ray spectrum, thus confirming the results obtained in earlier studies \citep[][]{2007MNRAS.382L..82G,2007ApJ...665..980T,2009AdSpR..43.1036F}.

\subsection{Jet Emission and the Soft X-ray Deficit}\label{subsec:jet_dscsn}
\begin{figure*}
\begin{center}
\hbox{
      \includegraphics[width=8cm]{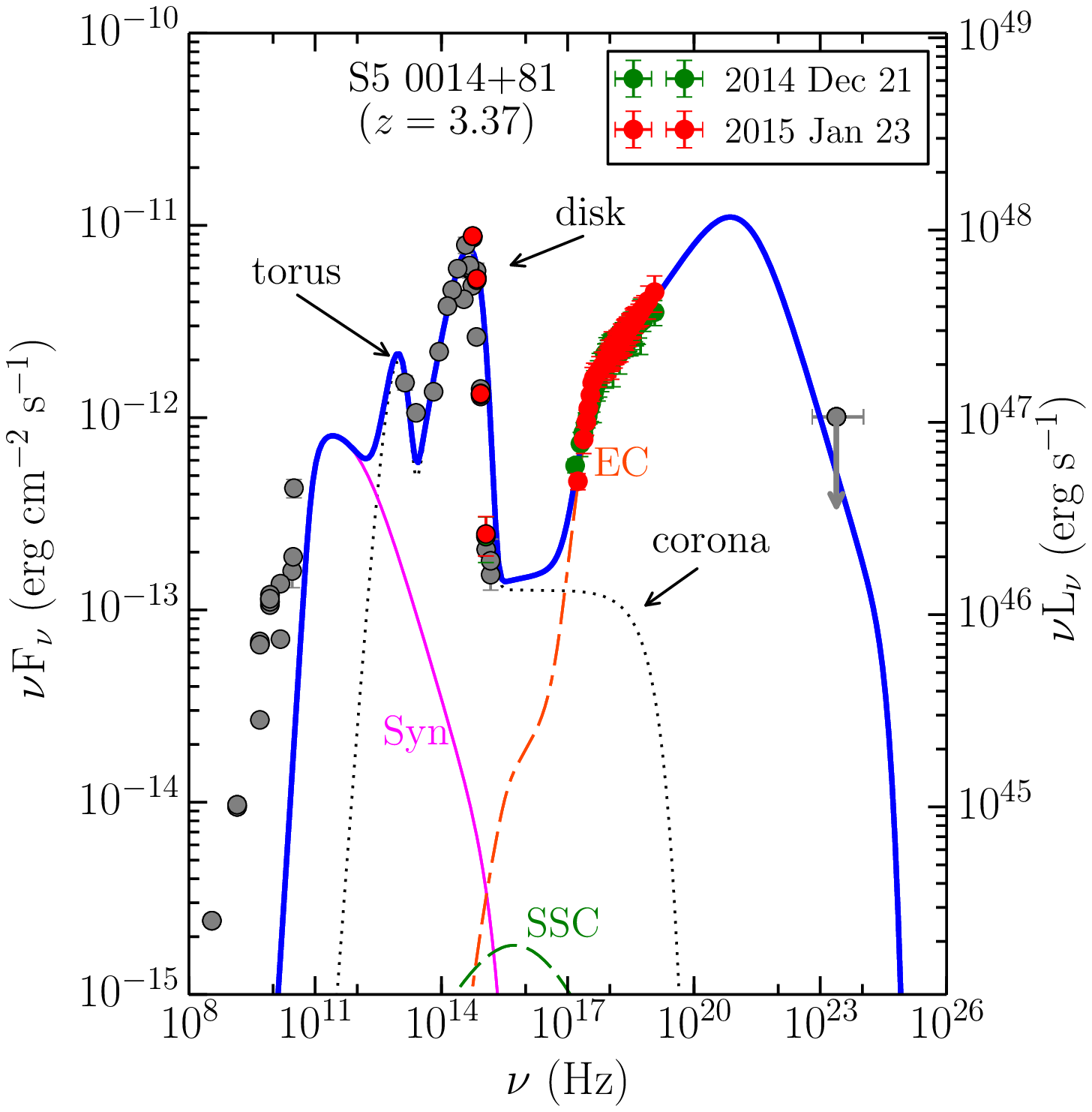}
      \includegraphics[width=8cm]{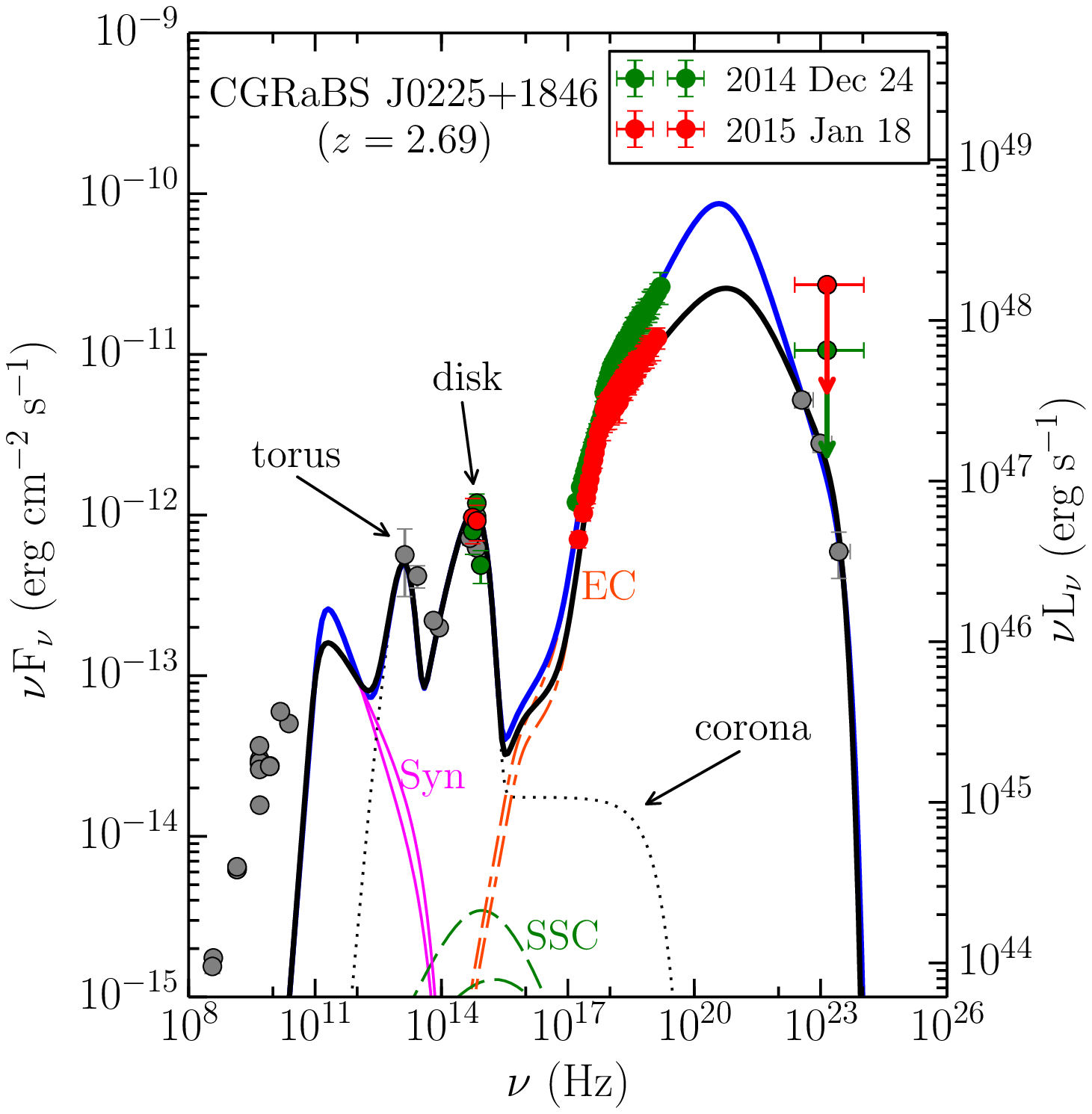}
     }
\hbox{
      \includegraphics[width=8cm]{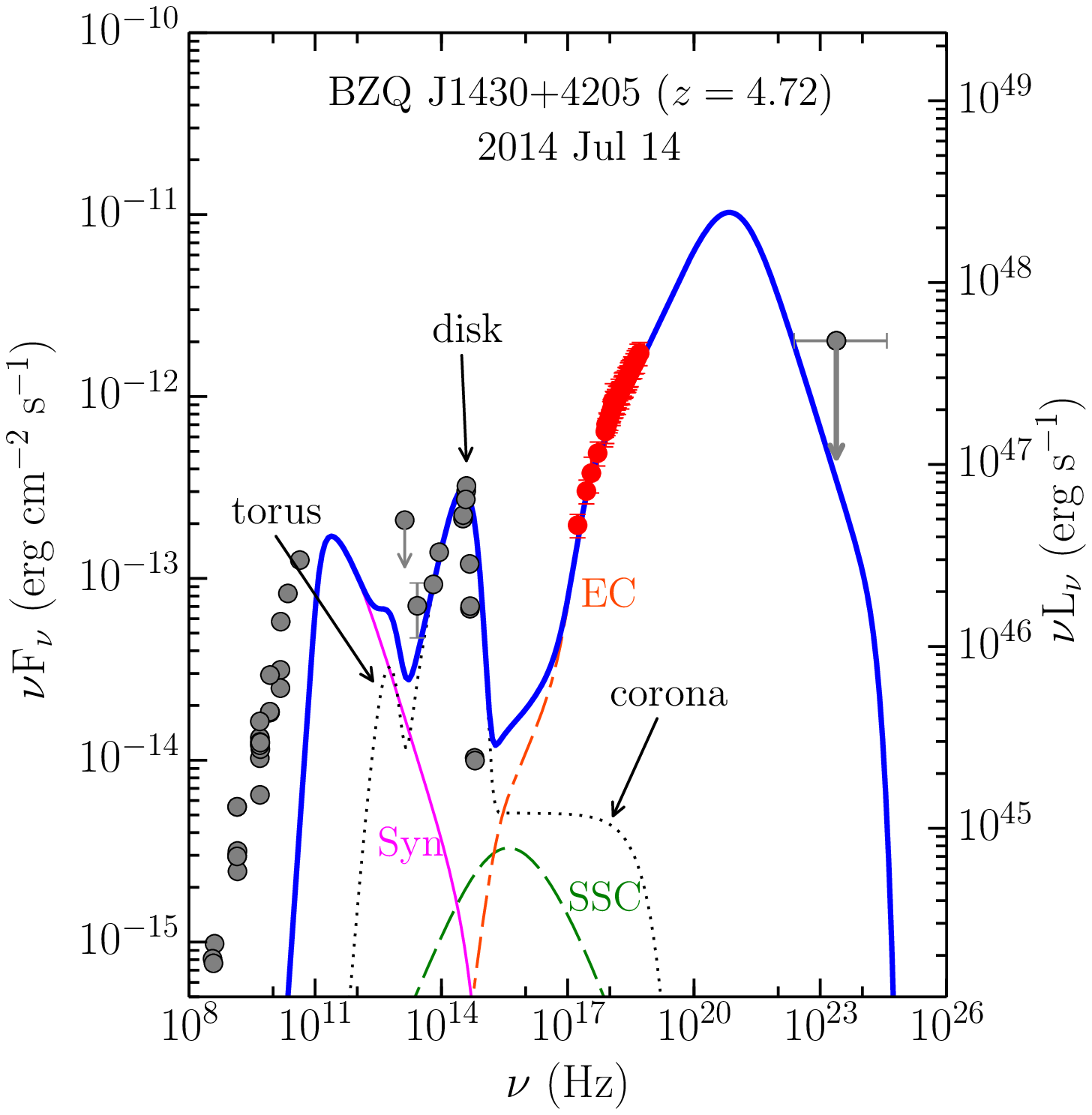}
      \includegraphics[width=8cm]{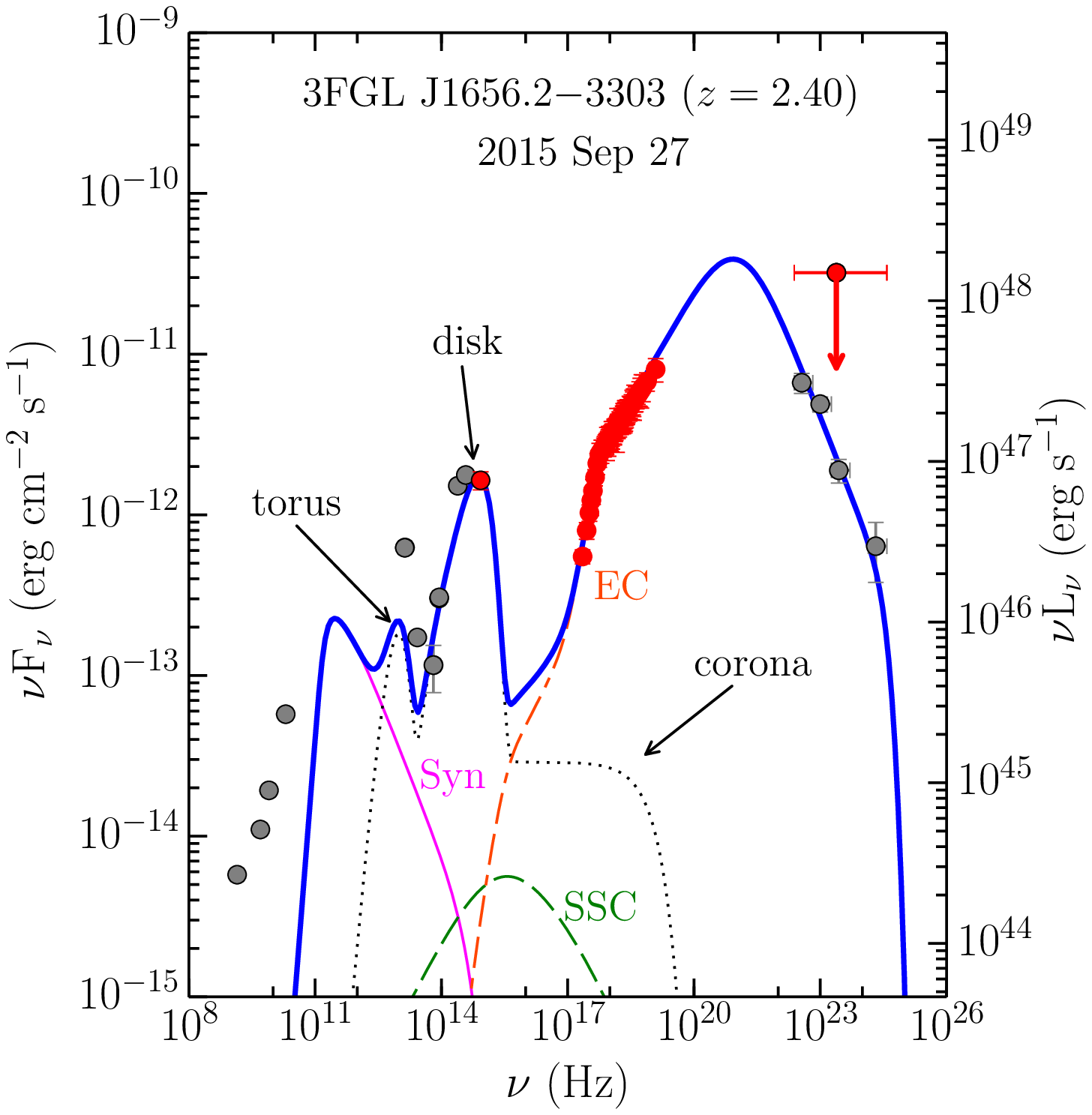}
     }
\caption{Spectral energy distributions of high redshift blazars. Contemporaneous data from \swift~and \nustar~are shown with colored (other than grey) circles and non-simultaneous observations are represented with grey circles. \fermi-LAT grey circles correspond to $\sim$7.5 years average \gm-ray spectrum, whereas, downward arrows are 2$\sigma$ upper limits. Thermal emission from the torus, the accretion disk, and the X-ray corona is represented by black dotted line. Pink thin solid, green dashed, and orange dash-dash-dot lines correspond to synchrotron, SSC, and EC emission, respectively. Blue and black thick solid lines are the sum of all the radiative components.}\label{fig:sed_fit}
\end{center}
\end{figure*}

The broadband SEDs of all the four blazars can be well reproduced using a simple one zone leptonic emission model (Figure \ref{fig:sed_fit}). Following the blazar sequence, their synchrotron peak lies at sub-mm or at self-absorbed frequencies and this shifting leaves the accretion disk spectrum visible at optical-UV energies. By modeling the optical-UV spectrum with a standard accretion disk model, we constrained both the black hole masses and the accretion luminosities in all the four sources. All the objects are found to host more than a billion solar mass black holes at their centers with the central black hole mass of J0014+81 is estimated as $\sim$10 billion solar mass, thus making it one of the most massive black holes ever found in radio-loud quasars beyond redshift $z=3$. For a consistency check, we compare the black hole masses obtained from SED modeling approach with that from single epoch optical spectroscopic line information. \citet[][]{1987SvA....31..136V} reported the following relation using the optical spectroscopy of J0014+81.
\begin{equation}
\frac{M_{\rm BH}}{R_{\rm BLR}}\approx1.3\times10^{10}~(\frac{M_{\odot}}{{\rm pc}})
\end{equation}

Following \citet[][]{2009MNRAS.397..985G}, we find the BLR radius as $\sim$1 parsec (see Table \ref{tab:sed_param}). This suggests $M_{\rm BH}\approx1.3\times10^{10}~M_{\odot}$ which is similar to that obtained from accretion disk modeling. For J1430+4205, we use C~{\sc iv} line and continuum parameters from the Sloan Digital Sky Survey data archive and adopt the empirical relations of \citet{2011ApJS..194...45S}. This gives the central black hole mass as $\sim$1.7$\times$10$^{9}~M_{\odot}$ which agrees within a factor of two to that derived by SED modeling approach. We consider the published C~{\sc iv} line parameters from \citet[][]{2008A&A...480..715M} and the derived black hole mass of J1656$-$3303 is $\sim$6.3 $\times$ 10$^{8}~M_{\odot}$ which reasonably matches within a factor of three to that obtained from modeling. On the other hand, we could not get any spectroscopic line information in literature for J0225+1846. Furthermore, the accretion disk is also found to be extremely luminous in all the sources with $L_{\rm disk}>10^{46}$ \lum. The level of synchrotron emission is constrained from the archival radio data and is kept low enough so as not to overproduce the observed IR-optical-UV SED. Another deciding factor is that we needed enough injected electron power to reproduce the high energy X-ray to \gm-ray SED via IC mechanism. Furthermore, in all the sources, the 0.3$-$79 keV X-ray spectra are very hard and the associated \gm-ray spectra are steeply falling, as expected from high redshift blazars. We interpret the entire X-ray to \gm-ray SED as a result of EC mechanism with the BLR as a primary reservoir of the seed photons. This sets the location of the emission region within the BLR in all the four objects, a feature generally seen in many high redshift FSRQs \citep[e.g.,][]{2010MNRAS.405..387G}. We constrain the spectral indices of the underlying electron population by reproducing the X-ray spectra and also by keeping into consideration the upper limits observed at LAT energies. Moreover, though not used for modeling, the long time average \fermi-LAT spectra assist us in having an idea about the typical shape of the falling part of the EC process.

In our sample of four blazars, two of them, J0014+81 and J0225+1846, were contemporaneously observed by \nustar~and \swift~in 2014 December and 2015 January. This, more than one episodes of monitoring, enables us to study the possible variations in their SEDs. As can be seen in Figure \ref{fig:sed_fit}, the SED of J0014+81 does not show any variation between these two epochs and, in fact, is similar to archival observations. On the other hand, J0225+1846 became faint in 2015 January compared to 2014 December. It should be noted that the slopes of the falling EC spectrum in both the activities states are primarily decided by the \fermi-LAT upper limits (and with \nustar~observations which controls the EC peak), but we also keep in mind its typical \gm-ray spectral shape as revealed by long time average LAT spectrum (Figure \ref{fig:sed_fit}). An interesting observation is the detection of low value of the \gm-ray flux upper limit at the time of the high X-ray state (2014 December, see also, Figure \ref{fig:fermi_lc}), which makes the falling EC spectrum to be steeper (Figure \ref{fig:sed_fit}). In 2015 January observation, on the other hand, J0225+1846 was in a relatively low X-ray state but the obtained upper limit indicates it to be slightly brighter in \gm-rays, thus making the EC spectrum flatter at \gm-ray energies. Now, on comparing these two SEDs and also noting that the \nustar~spectrum is harder in brighter state, we find that during the high activity phase, the EC peak shifted to lower frequencies, in accordance with the traditional blazar sequence.

\begin{figure*}
\begin{center}
\hbox{\hspace{-1.0cm}
      \includegraphics[width=9cm]{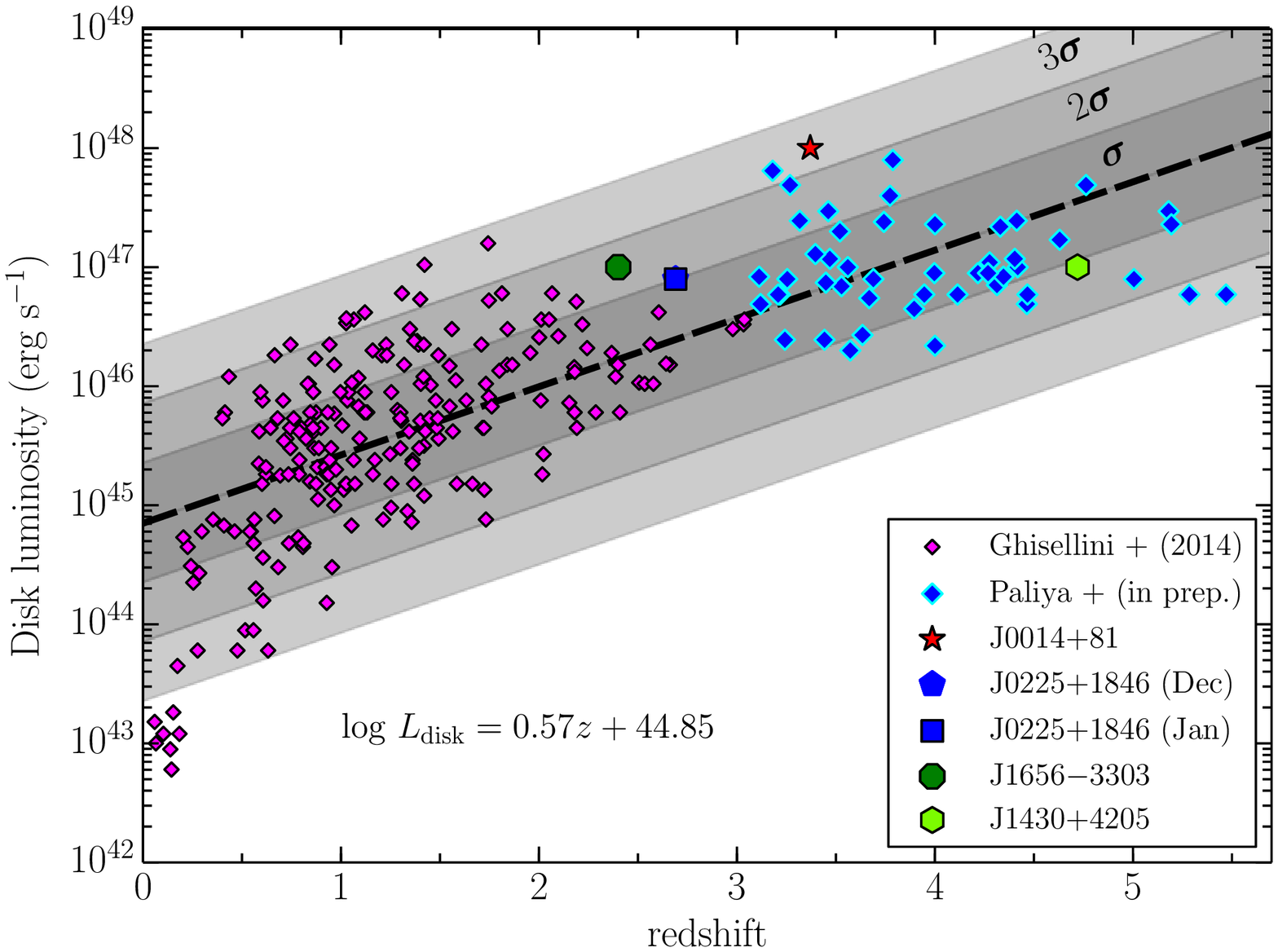}
      \includegraphics[width=9cm]{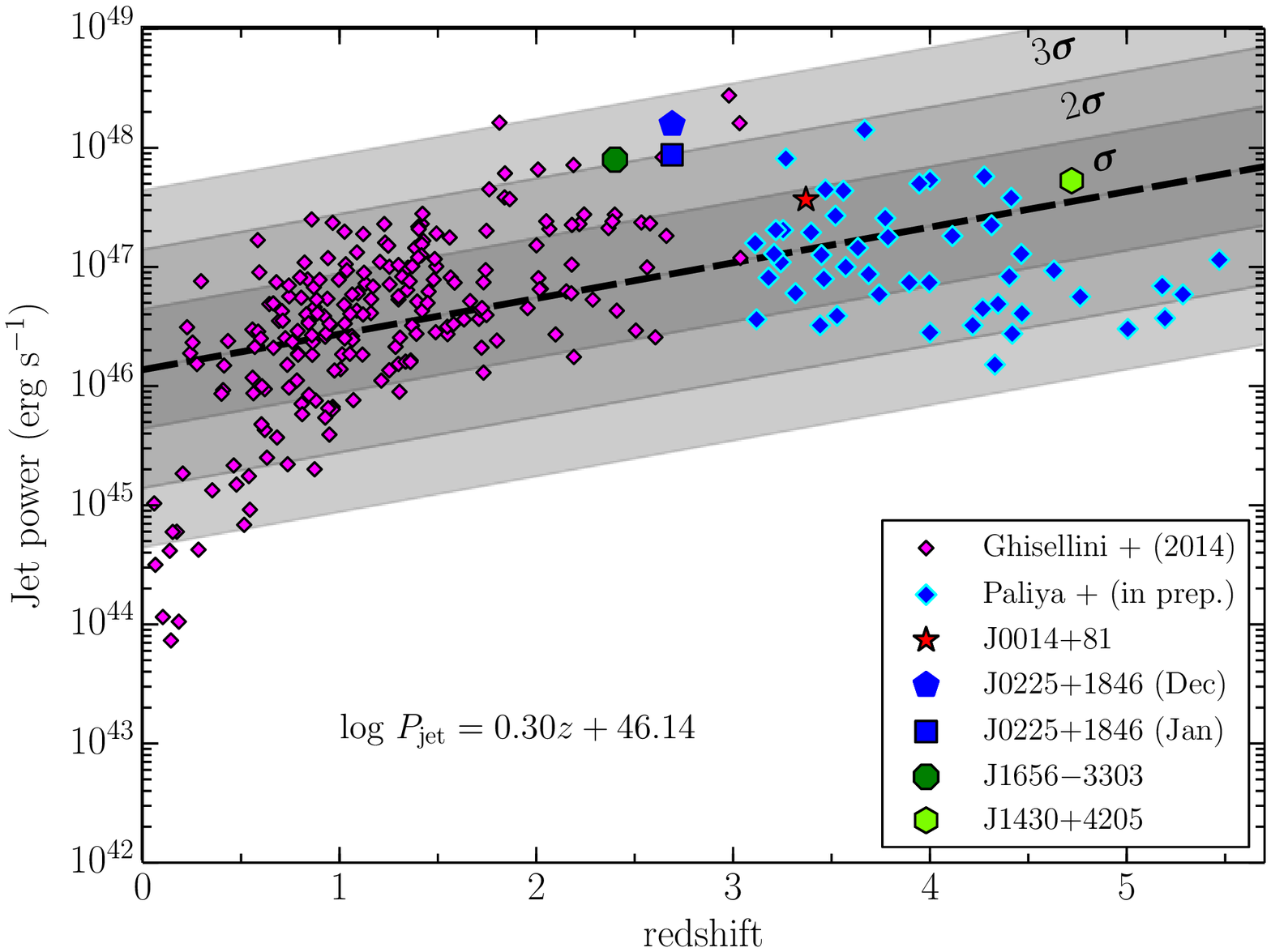}
     }
\hbox{\hspace{-1.0cm}
      \includegraphics[width=9cm]{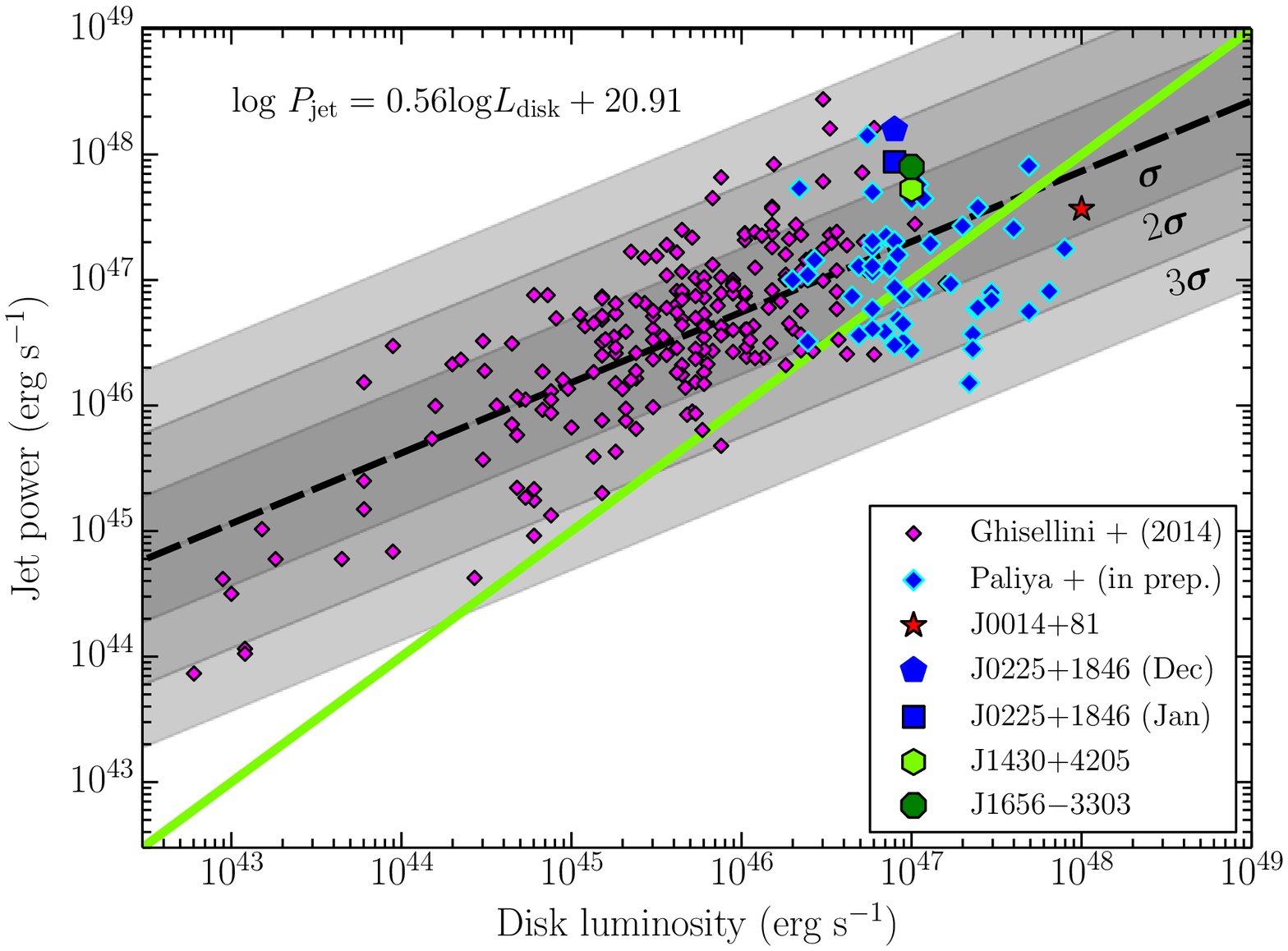}
      \includegraphics[width=9cm]{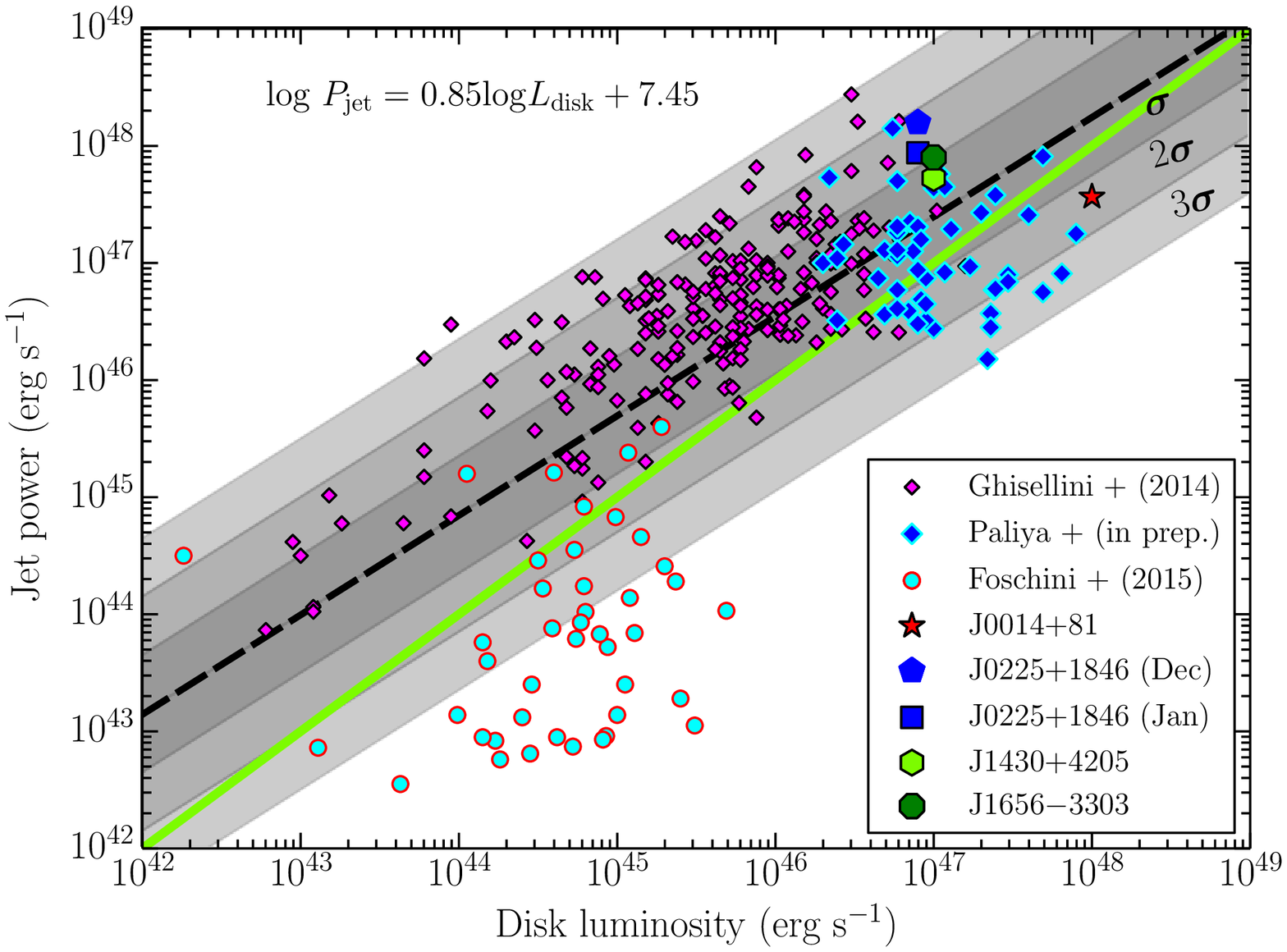}
     }
\caption{Top: Cosmic evolution of the accretion disk luminosity and the jet power of blazars upto redshift 5.47. Bottom: The accretion disk luminosity versus jet power for the known blazars (left) and when including RL-NLSy1 galaxies (right). The pink diamonds represent the blazars studied by \citet[][]{2014Natur.515..376G} and blue ones are modeled by us. The cyan circles are associated with RL-NLSy1 galaxies studied in \citet[][]{2015A&A...575A..13F}. The four high redshift blazars studied in this work are also shown. The black dashed lines correspond to the derived best linear fit, whereas, limegreen solid line in the bottom panels refers to the one-to-one correlation of the plotted quantities. The shaded areas represent $\sigma$, 2$\sigma$, and 3$\sigma$ dispersions with $\sigma=0.5$ dex. }\label{fig:disk_jet}
\end{center}
\end{figure*}

It is interesting to compare the physical properties of these four high redshift blazars with other blazars at similar redshifts and also with comparatively nearby objects. The main motivation here is to search for any possible trend or evolution of the physical parameters, such as the $L_{\rm disk}$ and total jet power ($P_{\rm jet}$) over cosmic history of time and also to check the relative position of the four sources studied here. With this in mind, we collect $L_{\rm disk}$ and $P_{\rm jet}$ values of all the 226 objects studied by \citet[][]{2014Natur.515..376G}, covering up to redshift 3.04. For higher redshift objects, we choose those sources from BZCAT \citep[][]{2015Ap&SS.357...75M} that have $z>$3.04, exhibit a hard X-ray spectrum \citep[X-ray photon index $<$1.7,see, e.g.,][]{2011ApJ...738...53S}, and are highly radio-loud\footnote{The selection of the sources from BZCAT is the part of our ongoing investigation of the physical properties of high redshift quasars and the details of the analysis will be presented in a forthcoming publication.} \citep[radio-loudness factor $R>$100, e.g.,][]{1989AJ.....98.1195K}. We then perform the broadband SED modeling of all these objects using the same leptonic emission model. We plot the variation of $L_{\rm disk}$ and $P_{\rm jet}$ with respect to the redshift and show them in the top panels of Figure \ref{fig:disk_jet}. We also present a plot between $L_{\rm disk}$ and $P_{\rm jet}$ in the bottom left panel of Figure \ref{fig:disk_jet}. Moreover, we over plot all the four sources considered in this work (including both the activity states of J0225+1846). We find a positive correlation between $L_{\rm disk}$ and $P_{\rm jet}$ with redshift, respectively, and also between themselves (Figure \ref{fig:disk_jet}). This is not unexpected because at higher redshifts, only the most powerful objects are visible. However, there are few points worth noticing. First, $L_{\rm disk}$ increases upto a redshift of $\sim$3$-$3.5 and after that it appears to saturate around 10$^{47}$ \lum. Second, the similar behavior is visible in $P_{\rm jet}$ versus redshift plot, where beyond redshift 4, $P_{\rm jet}$ tend to deviate more towards the lower power from the best linear fit. A comparison of the slopes obtained from the best linear fitting of $L_{\rm disk}$ and $P_{\rm jet}$ versus redshift indicates that $P_{\rm jet}$ does not increases at the same rate as $L_{\rm disk}$. In fact, the visual inspection suggests a relatively decrease in $P_{\rm jet}$ for $z>$3 objects. These findings are further confirmed in $L_{\rm disk}$ versus $P_{\rm jet}$ plot, where slope of the obtained linear fit (0.56 $\pm$ 0.03) is steeper than that derived by \citet[][]{2014Natur.515..376G}. As can be seen in Figure \ref{fig:disk_jet}, the change of slopes has occurred due to high redshift ($z>$3) blazars to occupy high $L_{\rm disk}$ and relatively low $P_{\rm jet}$ regime. In other words, though $P_{\rm jet}>L_{\rm disk}$ is true for relatively low redshift blazars \citep[][]{2014Natur.515..376G} but the same may not hold for blazars beyond redshift 3 or 4. However, we caution that a strong claim cannot be made because of two reasons: (i) blazars beyond redshift 4 are primarily discovered in IR-optical surveys \citep[e.g.,][]{2014ApJ...795L..29Y} and thus are relatively bright in the optical band, implying more luminous accretion disk, even though having moderate power jets, and (ii) it is difficult to precisely measure the jet power in blazars beyond redshift 3 due to lack of good quality high energy X-ray and \gm-ray observations. None of the sources, beyond redshift 3.04, are present in 3FGL catalog and only few of them are having sensitive X-ray data from facilities like \nustar. This could be the reason of the observed large scatter in $P_{\rm jet}$ for blazars beyond redshift 3. Observations from \nustar~are crucial to confirm/reject the trend seen in Figure \ref{fig:disk_jet}. On a different note, it is also of great interest to consider another class of AGN with powerful relativistic jets, the radio-loud narrow line Seyfert 1 (RL-NLSy1) galaxies. The detection of significant \gm-ray emission from some of these RL-NLSy1 galaxies have confirmed the idea that these sources too host powerful relativistic jets similar to blazars \citep[see, e.g.,][for a review]{2012nsgq.confE..10F}. We collect the $L_{\rm disk}$ and $P_{\rm jet}$ information for a sample of RL-NLSy1 galaxies studied by \citet[][]{2015A&A...575A..13F} and plot them along with the known blazars in the bottom right panel of Figure \ref{fig:disk_jet}. As can be seen, the RL-NLSy1 galaxies do not follow the $L_{\rm disk}-P_{\rm jet}$ correlation seen among blazars and occupy a distinct region where there jet power is considerably lesser than blazars and the disk luminosity is comparable to low jet power blazars. These objects are known to have a high accretion rate \citep[e.g.,][]{2006ApJS..166..128Z} but host a relatively low power jet, possibly due to harboring low mass black holes\footnote{However, once normalized by the black holes mass, the jet powers of RL-NLSy1 galaxies are consistent with blazars, indicating the scalability of the jet \citep[][]{2015A&A...575A..13F}.}. Considering the four sources studied here, we find them to occupy a place in  $L_{\rm disk}$ versus $P_{\rm jet}$ plot where their jet power exceeds the accretion luminosities, except for J0014+81. Furthermore, in the redshift evolution diagram, they appear to follow the common trend with no major exception.

\begin{figure*}
\hbox{
      \includegraphics[width=9cm]{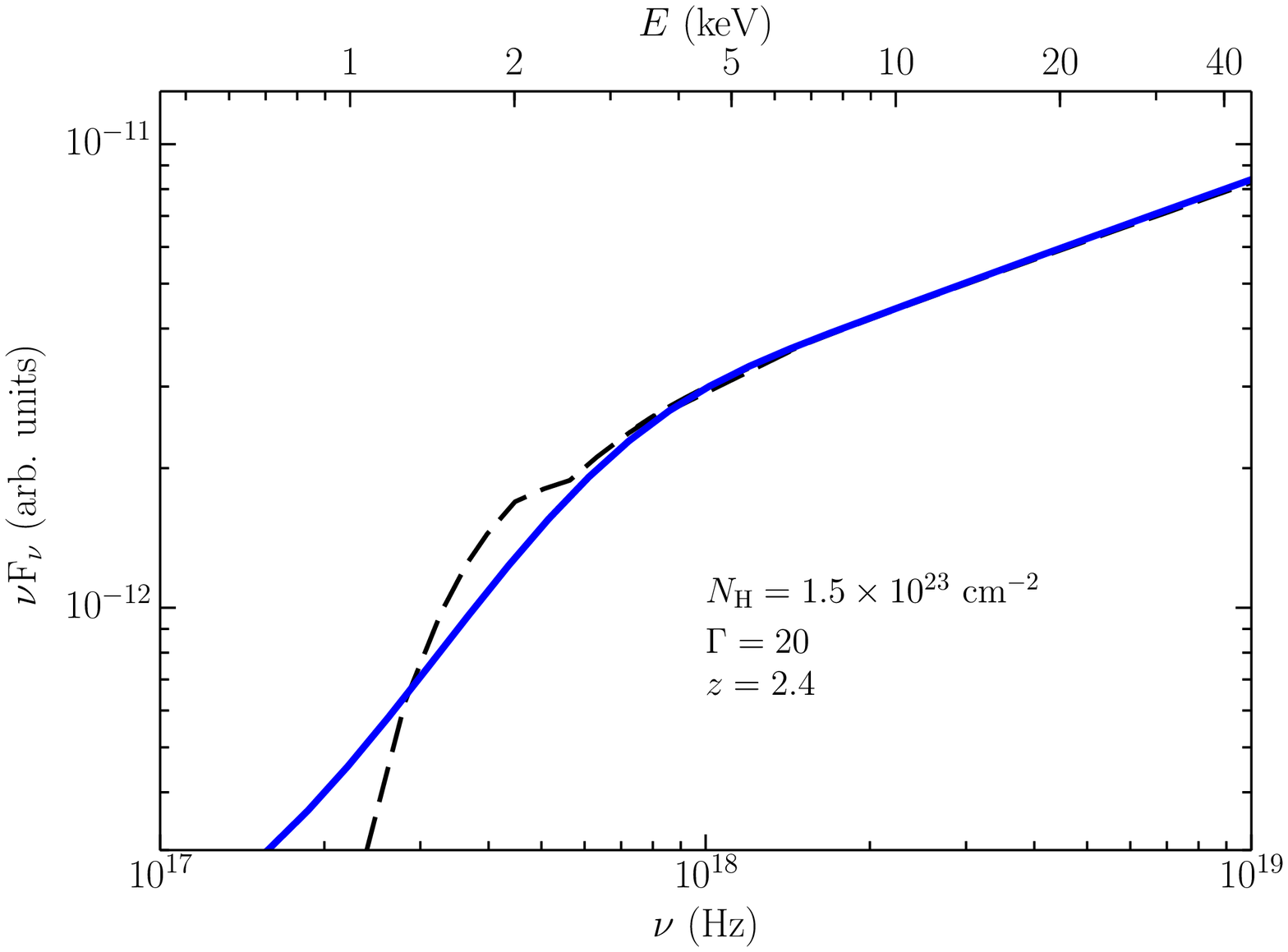}
      \includegraphics[width=9cm]{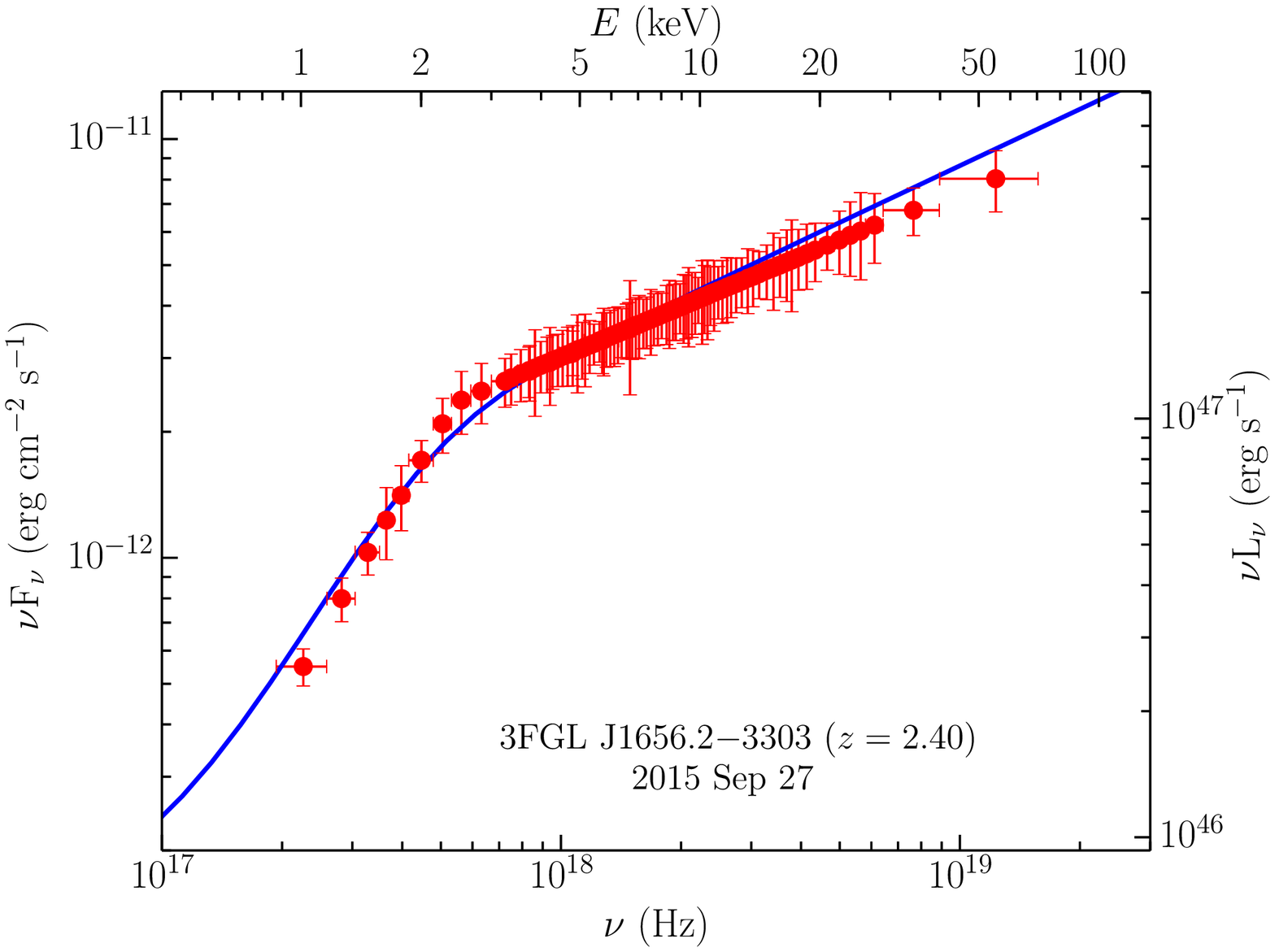}
}
\caption{Left: Comparison of the EC spectrum (blue solid line) with a power law (photon index 1.6) spectrum with absorption intrinsic to the quasar frame (black dashed line). The bulk Lorentz factor is assumed as $\Gamma=20$ to match the absorbed power law spectrum and also we adopt $\gamma_{\rm min}=1$. The spectra have been arbitrarily normalized. Right: The zoomed version of the X-ray SED of J1656$-$3303.}\label{fig:PL_EC}
\end{figure*}

The soft X-ray deficit observed in many jetted sources can also be reproduced by the intrinsic curvature of the EC emission from the jet \citep[see, e.g.,][]{2007MNRAS.382L..82G,2007ApJ...665..980T,2007ApJ...669..884S,2008MNRAS.386..945T,2009AdSpR..43.1036F}. In fact, the SED model used to reproduce the broadband spectrum of a high redshift blazar do predicts a smooth flattening of the X-ray spectrum below few keV, although the accurate measurement of the shape and the location of the break depends on the SED parameters and also on the ambient photon distribution \citep[e.g.,][]{2007ApJ...665..980T}. Below the break, the EC slope reflects the shape of the seed photon distribution ($F(\nu)\propto\nu^2$ for a blackbody distribution). To emphasize more, we plot an absorbed power law spectrum with redshifted column density $N_{\rm H}=1.5\times10^{23}$ cm$^{-2}$ and an EC spectrum, for a source located at a redshift of 2.4 and moving with $\Gamma=20$, in Figure \ref{fig:PL_EC}. As discussed above, a break in the EC spectrum is visible at a frequency of $\nu_{\rm break}\simeq\nu_{\rm seed}\Gamma^2\gamma_{\rm min}^2/(1+z)$, where $\nu_{\rm seed}$ is the peak frequency of the ambient photon field providing seed photons for IC scattering. Assuming BLR photon field as a predominant source of seed photons ($\nu_{\rm seed}=\nu_{\rm Ly-\alpha}$) and fixing the minimum energy of the underlying electron population to $\gamma_{\rm min}=1$, we find that the break frequency depends only on the bulk Lorentz factor. The shapes of both the spectra are quite similar down to $\sim$1$-$2 keV (which will be lower for higher redshift objects), as can be seen in Figure \ref{fig:PL_EC}. Below this energy, EC asymptotically follows a spectrum $\propto\nu^2$ (reflecting the slope of the ambient photons), while intrinsically absorbed power law drops exponentially. Now, to compare these theoretical arguments with the observations, we plot the zoomed version of the X-ray part of the modeled SED of J1656$-$3303 in Figure \ref{fig:PL_EC}. Both the models, discussed above, can reasonably fit the data. However, based on our study (see Section \ref{subsec:X_ray_dscsn}), the hypothesis of host galaxy/warm absorption can be rejected. This leaves the presence of intrinsic curvature in the jet emission as the most plausible explanation of the soft X-ray deficit in the high redshift blazars and confirms the results of the earlier studies \citep[e.g.,][]{2007MNRAS.382L..82G,2007ApJ...665..980T,2009AdSpR..43.1036F}.

\section{Summary}\label{sec:summary}
 In this paper, we study the broadband physical properties of four high redshift blazars namely, J0014+81, J0225+1846, J1430+4205, and J1656$-$3303. We summarize our findings below.
\begin{enumerate}
\item A statistically significant \gm-ray emission has been detected from J0225+1846, confirming earlier predictions of this object as a \gm-ray emitter.
\item The broadband SEDs of all the four blazars are typical of their high redshift counterparts with optical-UV emission is dominated by the accretion disk radiation and high energy X-ray to \gm-ray spectra are dominated by jet emission processes.
\item All the sources are found to host more than a billion solar mass black holes at their centers and their accretion disk luminosity exceeds 10$^{46}$ \lum. Moreover, they occupy the high end of the $L_{\rm disk}$-$P_{\rm jet}$ correlation.
\item A detailed investigation of the joint XRT/\nustar~spectral fitting (and also RGS spectral fitting of J0225+1846) favors the jet based origin of the observed soft X-ray flattening in high redshift blazars rather than due to external effects, such as host galaxy absorption and/or warm absorber. 
\item Overall, the blazars beyond redshift 3 are tend to deviate from the known one-to-one accretion-jet correlation, however, further observations from facilities, e.g., \nustar, are necessary to confirm/reject this hypothesis.
\end{enumerate}

\acknowledgments
We are grateful to the referee for constructive comments on the manuscript. ACF thanks Greg Madejski for discussions on the curvature of blazar X-ray spectra and acknowledges support from ERC Advanced Grant 340442. This research has made use of data, software and/or web tools obtained from NASA’s High Energy Astrophysics Science Archive Research Center (HEASARC), a service of Goddard Space Flight Center and the Smithsonian Astrophysical Observatory. Part of this work is based on archival data, software, or online services provided by the ASI Science Data Center (ASDC). This research has made use of the XRT Data Analysis Software (XRTDAS) developed under the responsibility of the ASDC, Italy. This research has also made use of the NuSTAR Data Analysis Software (NuSTARDAS) jointly developed by the ASI Science Data Center (ASDC, Italy) and the California Institute of Technology (Caltech, USA). This research has made use of the NASA/IPAC Extragalactic Database (NED) which is operated by the Jet Propulsion Laboratory, California Institute of Technology, under contract with the National Aeronautics and Space Administration.
Funding for SDSS-III has been provided by the Alfred P. Sloan Foundation, the Participating Institutions, the National Science Foundation, and the U.S. Department of Energy Office of Science. The SDSS-III web site is http://www.sdss3.org/.

SDSS-III is managed by the Astrophysical Research Consortium for the Participating Institutions of the SDSS-III Collaboration including the University of Arizona, the Brazilian Participation Group, Brookhaven National Laboratory, Carnegie Mellon University, University of Florida, the
French Participation Group, the German Participation Group, Harvard University, the Instituto de Astrofisica de Canarias, the Michigan State/Notre Dame/JINA Participation Group, Johns Hopkins University, Lawrence Berkeley National Laboratory, Max Planck Institute for Astrophysics, Max Planck Institute for Extraterrestrial Physics, New Mexico State University, New York University, Ohio State University, Pennsylvania State University, University of Portsmouth, Princeton University, the Spanish Participation Group, University of Tokyo, University of Utah, Vanderbilt University, University of Virginia, University of Washington, and Yale University. 

\bibliographystyle{apj}
\bibliography{Master}

\begin{table}
\caption{Results of the LAT data analysis of high redshift blazars studied in this work, covering the first $\sim$7.5 years of \fermi~operation. Col. [1]: Object name; Col.[2]: model used (PL:
power law, LP: logParabola); Col.[3]: integrated $\gamma$-ray flux (0.1$-$300 GeV), in units of 10$^{-8}$
\phflux, UL corresponds to the 2$\sigma$ upper limit; Col.[4] and [5]: spectral parameters (see definitions in the text); Col.[6]: test statistic; and Col.[7]: curvature of test statistic}\label{tab:avg_gamma}
\begin{center}
\begin{tabular}{lcccccc}
\hline
Name & Model & $F_{0.1-300~{\rm GeV}}$ & $\Gamma_{0.1-300~{\rm GeV}}/\alpha$ & $\beta$ & TS & TS$_{\rm curve}$\\
~[1] & [2] & [3] & [4] & [5] & [6] & [7] \\
\hline
J0014+81    & PL & 0.15 (UL) & --- &                 & 1.47  &     \\
            & LP & ---       & --- & ---             & ---   & --- \\
J0225+1846  & PL & 2.65 $\pm$ 0.21 & 2.99 $\pm$ 0.09 &                 & 188.43 & \\
            & LP & 2.49 $\pm$ 0.25 & 2.86 $\pm$ 0.14 & 0.45 $\pm$ 0.21 & 195.51 & 9.82 \\
J1430+4205  & PL & 0.42 (UL) & --- &                 & 5.62  &     \\
            & LP & ---       & --- & ---             & ---   & --- \\
J1656$-$3303& PL & 4.37 $\pm$ 0.32 & 2.87 $\pm$ 0.07 &                 & 207.21  & \\
            & LP & 4.37 $\pm$ 0.31 & 2.87 $\pm$ 0.06 & 0.00 $\pm$ 0.00 & 207.21  & 0.00 \\
\hline
\end{tabular}
\end{center}
\end{table}

\begin{table}
\caption{Fit parameters from the joint \nustar --XRT spectra. Column densities are in units of $10^{22}$~cm$^{-2}$.}
\label{table_xrtnustar}
\begin{center}
\begin{tabular}{l  l l l l l l}
\hline
\hline
Source & Model  & $\Gamma$ & $N_\textrm{H}$ & $E_\textrm{break}$ & $\Gamma_2$ & $\chi^2_\nu/$d.o.f.\\
\hline
J0014+81	& M1 	&   $1.60\pm0.01$ & &&&  1.19/206\\
		 	& M2 	&   $1.67\pm0.02$ & $4\pm1$ & && 1.03/205\\
		 	& M3	&   $1.2\pm0.1$ & & $2.3_{-0.4}^{+0.7}$ & $1.70\pm0.02$ & 0.97/204\\

\hline
J0225+1846 	\\
(60001101002)& M1 	&  $1.40\pm0.01$ & &&&  1.84/667\\
			& M2 	&  $1.51\pm0.01$ & $6.4\pm1$ & && 1.15/675\\
		 	& M3	&  $1.00\pm0.05$ & & $4.3\pm0.4$ & $1.54\pm0.02$ & 1.03/673\\
(60001101004)& M1 	&  $1.47\pm0.01$ & &&&  1.84/667$^{1}$\\
		 	& M2	&  $1.65\pm0.02$ & $13\pm2$ & && 1.15/675$^{1}$\\
		 	& M3	&  $1.01_{-0.06}^{+0.04}$ & & $4.5_{-0.7}^{+0.5}$ & $1.67\pm0.03$& 1.03/673$^{1}$\\
\hline
J1430+4205	& M1 	&   $1.45\pm0.02$ & &&&  1.20/78\\
		 	& M2 	&   $1.51\pm0.03$ & $7_{-2}^{+3}$ & && 1.10/77\\
		 	& M3	&   $1.3_{-0.4}^{+0.1}$ & & $5_{-3}^{+2}$ & $1.59_{-0.06}^{+0.08}$ & 1.03/76\\
\hline
J1656$-$3303 & M1 &  $1.42\pm0.02$ & & & & 1.63/95\\
			& M2 & $1.58\pm0.03$& $11\pm1$ & & & 1.03/94\\
			& M3 &  $0.8\pm0.2$ & & $2.2\pm0.3$ & $1.60\pm0.03$ & 0.97/93\\
\hline
\hline

\end{tabular}
\end{center}
$^{1}$The $\chi^2$ values are identical for the two observations of J0225+1486 because the spectra are fit simultaneously.
\end{table}

\begin{table}
\footnotesize
\caption{Fit parameters from the \xmm\ spectra. Column densities are in units of $10^{22}$~cm$^{-2}$.}
\label{table_xmm}
\begin{center}
\begin{tabular}{l  l l l l l l}
\hline
\hline
Source & Model  & $\Gamma$ & $N_\textrm{H}$ & $E_\textrm{break}$ & $\Gamma_2$ & $\chi^2_\nu/$d.o.f.\\
\hline
J0014+81	& M1 	&   $1.49\pm0.01$ & &&&  0.97/116\\

\hline
J0225+1846 	\\

(0150180101)& M1 	&  $1.292\pm0.003$ & &&&  3.29/1062\\
			& M2 	&  $1.401\pm0.004$ & $1.43\pm0.05$ & && 1.12/1059\\
		 	& M3	&  $0.83\pm0.02$ & & $1.04\pm0.02$ & $1.383\pm0.004$ & 1.12/1056\\
		 	
(0690900101)& M1 	&  $1.312\pm0.003$ & &&&   3.29/1062$^{1}$\\
		 	& M2	&  $1.425\pm0.005$ & $1.45\pm0.05$ & && 1.12/1059$^{1}$\\
		 	& M3	&  $0.93\pm0.02$ & & $1.18\pm0.03$ & $1.42\pm0.01$& 1.12/1056$^{1}$\\
		 	
(0690900201)& M1 	&  $1.139\pm0.005$ & &&&   3.29/1062$^{1}$\\
		 	& M2	&  $1.22\pm0.01$ & $1.3\pm0.1$ & && 1.12/1059$^{1}$\\
		 	& M3	&  $0.75\pm0.05$ & & $1.07\pm0.06$ & $1.21\pm0.01$& 1.12/1056$^{1}$\\
\hline
J1430+4205\\
(0111260101)& M1 	&  $1.70\pm0.07$ & &&&  1.06/150\\
			& M2 	&  $1.9\pm0.1$ & $2\pm1$ & && 0.80/147\\
		 	& M3	&  $<1.5$ & & $0.5_{-0.1}^{+0.4}$ & $1.8\pm0.1$ & 0.82/144\\
		 	
(0111260701)& M1 	&  $1.60\pm0.02$ & &&&   1.06/150$^{1}$\\
		 	& M2	&  $1.71\pm0.03$ & $1.9\pm0.4$ & && 0.80/147$^{1}$\\
		 	& M3	&  $1.1_{-1.0}^{+0.3}$ & & $0.7_{-0.2}^{+0.3}$ & $1.68_{-0.03}^{+0.05}$& 0.82/144$^{1}$\\
		 	
(0212480701)& M1 	&  $1.41\pm0.02$ & &&&   1.06/150$^{1}$\\
		 	& M2	&  $1.49\pm0.03$ & $1.7\pm0.5$ & && 0.80/147$^{1}$\\
		 	& M3	&  $<0.95$ & & $0.6\pm0.1$ & $1.46\pm0.02$& 0.82/144$^{1}$\\

\hline
J1656$-$3303 & M1 &  $1.12\pm0.01$ & & & & 1.08/130\\
			& M2 & $1.17\pm0.02$& $0.9\pm0.3$ & & & 0.99/129\\
			& M3 & $0.93\pm0.05$ & & $1.6\pm0.2$ & $1.18\pm0.02$ & 0.94/128\\
\hline
\hline

\end{tabular}
\end{center}
$^{1}$The $\chi^2$ values are identical for the three observations of J0225+1486 and J1430+4205 because the spectra are fit simultaneously. Additionally, for J0225+1486 source we include the RGS data and use \emph{tbnew} instead of \emph{tbabs}.
\end{table}

\begin{table*}
{\small
\begin{center}
\caption{List of the Parameters used in the SED modeling. Col.[1]: Name of the object; Col.[2]: date of observation; Col.[3]: mass of the central black hole, in log scale; Col.[4]: accretion disk luminosity, in \lum, in log scale; Col.[5] and [6]: spectral indices of the electron energy distribution; Col.[7]: magnetic field, in Gauss; Col.[8]: particle energy density, in erg cm$^{-3}$; Col.[9]: bulk Lorentz factor; Col.[10]: break Lorentz factor; Col.[11]: maximum Lorentz factor; Col.[12]: distance of the emission region from central black hole, in parsec; Col.[13]: Size of BLR, in parsec; and Col.[14]: characteristic temperature of the dusty torus, in Kelvin.}\label{tab:sed_param}
\begin{tabular}{lccccccccccccc}
\tableline
\tableline
Name & Date & $M_{\rm BH}$ & $L_{\rm disk}$ & $n1$ & $n2$ & $B$ & $U_{\rm e}$ & $\Gamma$ & $\gamma_{\rm b}$ & $\gamma_{\rm max}$ & $R_{\rm diss}$ & $R_{\rm BLR}$ & $T_{\rm torus}$ \\
~[1] & [2] & [3] & [4] & [5] & [6] & [7] & [8] & [9] & [10] & [11] & [12] & [13] & [14] \\
\tableline
J0014+81     & 2014 Dec 21 & 10.04 & 48.0 & 2.3 & 4.5 & 2.2 & 0.001 & 10 & 116 & 5000 & 0.89 & 1.02 & 500 \\
J0225+1846   & 2014 Dec 24 & 9.40  & 46.9 & 1.9 & 4.7 & 1.0 & 0.023 & 16 & 46  & 1000 & 0.26 & 0.29 & 580 \\
             & 2015 Jan 18 & 9.40  & 46.9 & 2.3 & 4.3 & 1.0 & 0.015 & 14 & 71  & 1000 & 0.26 & 0.29 & 580 \\
J1430+4205   & 2014 Jul 14 & 9.54  & 47.0 & 1.9 & 4.5 & 1.6 & 0.022 & 14 & 79  & 3500 & 0.20 & 0.31 & 400 \\
J1656$-$3303 & 2015 Sep 27 & 9.30  & 47.0 & 2.3 & 4.4 & 1.0 & 0.051 & 14 & 68  & 3500 & 0.15 & 0.32 & 400 \\
\tableline
\end{tabular}
\tablecomments{There are no differences in the SEDs of J0014+81 for 2014 Dec 21 and 2015 Jan 23 observations (see Figure \ref{fig:sed_fit}). Therefore, here we provide modeling parameters associated with 2014 Dec 21 SED only.}
\end{center}
}
\end{table*}

\begin{table*}
\begin{center}
\caption{Various jet powers (in \lum) derived from the SED modeling of the high redshift blazars studied in this work. Col.[1]: Name of the object; Col.[2]: date of observation; Col.[3], [4], [5], and [6]: jet powers in magnetic field, radiation, electrons, and protons, respectively, in log scale.}\label{tab:jet_power}
\begin{tabular}{lccccc}
\tableline
\tableline
Name & Date & $P_{\rm mag}$ & $P_{\rm rad}$ & $P_{\rm ele}$ & $P_{\rm kin}$ \\
~[1] & [2] & [3] & [4] & [5] & [6] \\ 
\tableline
J0014+81     & 2014 Dec 21 & 47.13 & 45.92 & 44.72 & 47.35 \\
J0225+1846   & 2014 Dec 24 & 45.80 & 46.76 & 45.57 & 48.18 \\
             & 2015 Jan 18 & 45.68 & 46.21 & 45.26 & 47.93 \\
J1430+4205   & 2014 Jul 14 & 45.86 & 46.30 & 45.20 & 47.70 \\           
J1656$-$3303 & 2015 Sep 27 & 45.21 & 46.21 & 45.32 & 47.89 \\          
\tableline
\end{tabular}
\end{center}
\end{table*}

\newpage

\newpage

\newpage

\newpage

\newpage

\end{document}